# Timed material self-assembly controlled by circadian clock proteins

*Gregor Leech[1], Lauren Melcher[2†], Michelle Chiu[3†], Maya Nugent[1], Lily Burton[4], Janet Kang[5], Soo Ji Kim[4], Sourav Roy[6], Leila Farhadi[6], Jennifer L. Ross[6], Moumita Das[2,7], Michael J. Rust[5], Rae M. Robertson-Anderson[1*]*

[1]Department of Physics and Biophysics, University of San Diego, San Diego, California 92110, United States

[2]School of Mathematical Sciences, Rochester Institute of Technology, Rochester, New York 14623, United States

[3]Graduate Program in Biophysical Sciences, University of Chicago, Chicago, Illinois 60637, United States

[4]Department of Biochemistry and Molecular Biophysics, University of Chicago, Chicago, Illinois 60637, United States

[5]Department of Molecular Genetics and Cell Biology and Department of Physics, University of Chicago, Chicago, Illinois 60637, United States

[6]Department of Physics, Syracuse University, Syracuse, New York 13244, United States

[7]School of Physics and Astronomy, Rochester Institute of Technology, Rochester, New York 14623, United States

[†]These authors contributed equally to this work.

**Abstract**

Active biological molecules present a powerful, yet largely untapped, opportunity to impart autonomous regulation to materials. Because these systems can function robustly to regulate when and where chemical reactions occur, they have the ability to bring complex, life-like behavior to synthetic materials. Here, we achieve this design feat by using functionalized circadian clock proteins, KaiB and KaiC, to engineer time-dependent crosslinking of colloids. The resulting material self-assembles with programmable kinetics, producing macroscopic changes in material properties, via molecular assembly of KaiB-KaiC complexes. We show that colloid crosslinking depends strictly on the phosphorylation state of KaiC, with kinetics that are synced with KaiB-KaiC complexing. Our microscopic image analyses and computational models indicate that the stability of colloidal super-structures depends sensitively on the number of Kai complexes per colloid connection. Consistent with our model predictions, a high concentration stabilizes the material against dissolution after a robust self-assembly phase, while a low concentration allows circadian oscillation of material structure. This work introduces the concept of harnessing biological timers to control synthetic materials; and, more generally, opens the door to using protein-based reaction networks to endow synthetic systems with life-like functional properties.

## Introduction

The current state-of-the-art in next-generation materials design is to create structures that can achieve desired functions in response to external perturbations, such as self-repair in response to damage. Looking beyond this stimulus-response framework, we envision autonomously functional materials that can not only respond directly to their environment, but also have the capability to store a memory of past events and dynamically change their own properties. Such materials could be used to create dynamic sequestration devices that filter toxins on a programmable schedule, or medical implants that self-assemble and restructure to protect and suture wounds and dissolve once fully healed.

An attractive strategy to equip materials with robust autonomous function is the use of distributed information processing throughout the material, rather than a central controller. This concept is similar to the function of interacting networks of biomolecules in living cells, which provide finely-tuned spatiotemporal regulation of physiology. In many cases, a small number of interacting network components can be isolated from the cell and retain modular function to achieve tasks such as defining structures with a specific size[1], generating spatial patterns[2], or keeping time[3]. On a larger scale, the collective action of these biomolecules provides a way for energy flux to impart non-equilibrium properties into structures to create active matter. The last two decades have seen tremendous progress on identifying and understanding the emergent properties of active matter from active colloids, to molecular motor-driven active biomaterials, to soft robotics and living concrete[4–12]. However, engineering autonomous materials with robust, kinetically-controlled activity, inherent to living systems, remains a grand challenge in materials science[13–15].

Here, we develop the proof-of-concept of an autonomous material with properties that are temporally programmed by biological signaling molecules using protein components derived from the cyanobacterial circadian clock (Fig. 1). In their natural context, KaiA, KaiB, and KaiC proteins generate a self-sustaining ~24-hour rhythm that is used to synchronize physiology with the external light-dark cycle. Remarkably, these proteins can be removed from their cellular context and can generate stable oscillations in a reconstituted in vitro system[3,16–18].

These oscillations can be detected as an ordered pattern of multisite phosphorylation on KaiC, which acts as a signaling hub that binds and releases protein partners throughout the cycle[19–21]. In brief, KaiC consists of two tandem ATPase domains, CI and CII, arranged into a hexameric ring (one blue circle in Fig. 1A represents one subunit consisting of a CI-CII pair). KaiA binds to the CII domain of KaiC, which stimulates autophosphorylation[22,23]. Phosphorylation accumulates slowly throughout the day, first on Thr432, then on Ser431. When Ser431 is heavily phosphorylated (shown in Fig. 1D), corresponding to dusk, ring-ring stacking interactions allow the CII domain to regulate the slow ATPase cycle in CI[24]. The post-hydrolysis state of CI allows KaiB to bind to KaiC[25] and six KaiB molecules to assemble cooperatively on the CI ring[26,27].

This nighttime KaiB-KaiC complex sequesters and inactivates KaiA, closing a negative feedback loop to inhibit further phosphorylation and allowing KaiC to dephosphorylate. Unphosphorylated KaiC then releases KaiB and is ready to begin the cycle again. The kinetics of the phosphorylation rhythms are remarkably robust to temperature and protein concentration, yet can be tuned dramatically by single amino acid substitutions in the Kai proteins[28,29]. The system is also remarkably thrifty in its energy consumption—with each KaiC molecule consuming only 15 ATP per day[30]. Thus, the Kai protein system is a uniquely attractive choice to develop into a synthetic tool to endow materials with programmable, autonomously time-dependent properties.

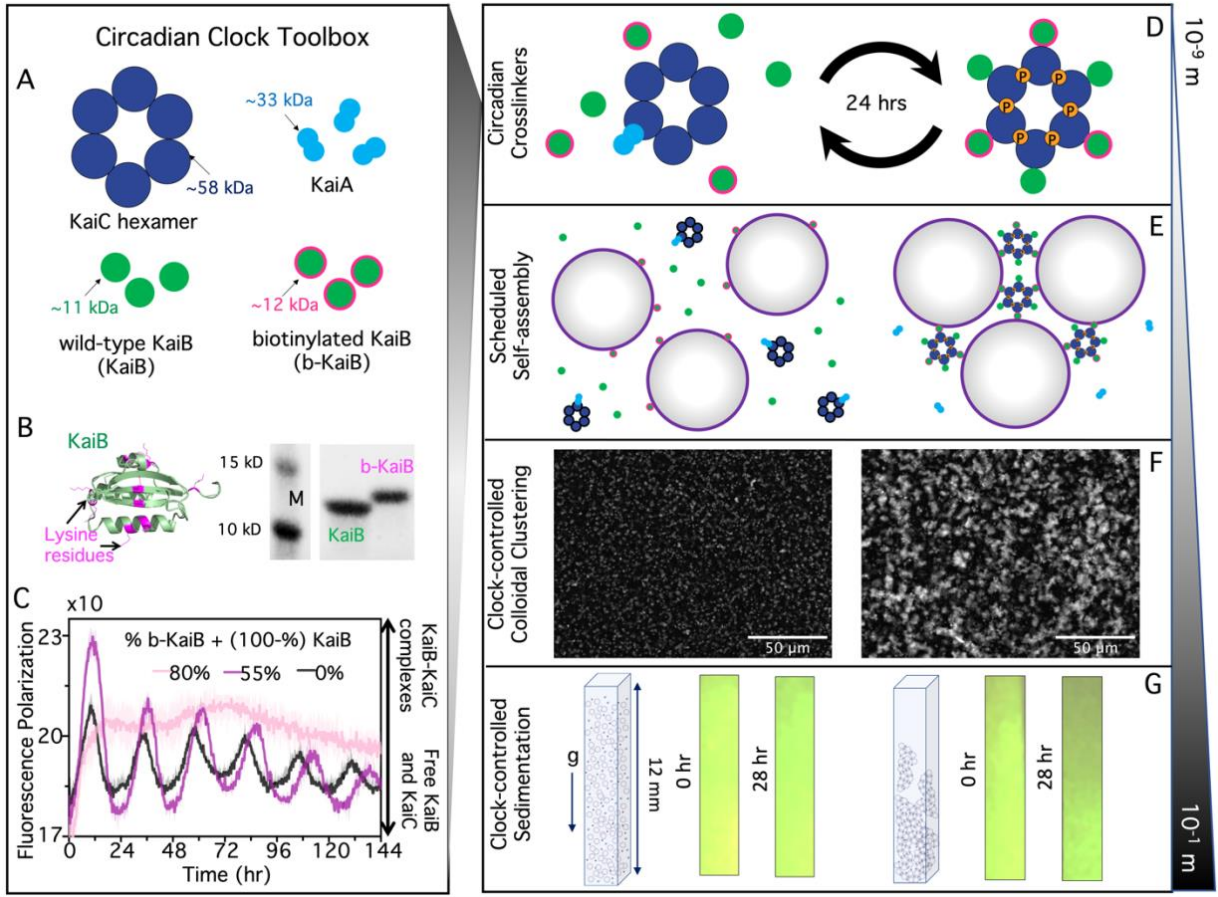

**Figure 1. Harnessing circadian clocks to engineer non-equilibrium materials across scales**. (**A**) We functionalize cyanobacteria clock proteins–hexameric KaiC rings (blue), KaiA dimers (cyan), and KaiB monomers (green)–to couple to materials by incorporating biotinylated KaiB (b-KaiB). (**B**) KaiB biotinylation: (*left*) potential sites of amine-reactive biotinylation (lysine residues in magenta) overlaid on the KaiB crystal structure (green)[31,32], (*right*) SDS-PAGE gel of unlabeled KaiB and biotinylated KaiB (b-KaiB), showing successful biotinylation indicated by a mobility shift of the biotinylated product (molecular weight standards [M] are 10 kDa and 15 kDa). (**C**) KaiABC reactions in the presence of biotinylated KaiB. Oscillations are measured by fluorescence polarization of FITC-labeled KaiB (0.2 µM), a read-out of KaiB-KaiC complex formation. All conditions contain 3.5 µM KaiB, with the specified fraction being b-KaiB. Oscillatory association of KaiB with KaiC is sustained with 55% b-KaiB (magenta), the percentage used in subsequent experiments, but not with 80% (pink). Each curve is an average of two replicates. (**D**) KaiB monomers bind cooperatively to KaiC rings in a phosphorylation-dependent manner (indicated by the orange 'P' circles), mediated by KaiA, and are subsequently released as KaiC dephosphorylates over a 24-hr cycle. We exploit the transition from free KaiB to KaiB fully assembled on a KaiC hexamer to create a time-dependent and phosphorylation-dependent change in crosslinking valency. (**E**) We incorporate the 'circadian crosslinkers' depicted in (D) into suspensions of 1-µm streptavidin-coated colloids to drive time-dependent crosslinking of colloids. (**F**) Microscope images of fluorescent streptavidin-coated colloids, mixed with KaiB, b-KaiB, and KaiC phosphorylation site mutants that cannot bind KaiB (left) or constitutively bind KaiB (right), show that KaiB-KaiC assembly selectively causes mesoscale clustering and connectivity of colloids. (**G**) Sedimentation of colloidal clock suspensions shown in (F) demonstrate pronounced settling of colloids after a day of incubation with the mutant that forms constitutively KaiB-KaiC complexes (right) compared to colloids mixed with the non-binding KaiC mutant (left).

## Results and Discussion

***Biotinylation of KaiB allows KaiC to selectively mediate material crosslinking.*** To harness the Kai protein system for materials activation, we aimed to exploit the changing oligomeric state of KaiB, induced by interaction with KaiC, throughout the circadian cycle mediated by KaiA. Namely, KaiB transitions from being free in solution to forming a hexameric KaiB-KaiC complex (KaiBC). We reasoned that functionalized KaiB molecules would not be effective crosslinkers of material components when KaiB is free in solution, but that assembly into the KaiBC complex might create a potent multivalent crosslinker where time-dependent KaiB-KaiC interactions can bridge multiple biotin-streptavidin bonds (Fig. 1D). To develop this tool and characterize its effect, we chose a commercial colloidal suspension as a model material platform (Fig. 1E). We hypothesized that as KaiBC complexes form over time, the number of colloids able to participate in KaiB-mediated crosslinks would increase, and we would observe a transition from a fluid-like suspension of single colloids, to a gel-like state with larger connected clusters of colloids (Fig. 1E-F). Consistent with this prediction, we expect macroscopic changes in the ability of the colloidal material to sediment (Fig. 1G).

To endow the KaiBC complex with time-dependent material crosslinking properties, we first needed to functionalize KaiB to bind to the colloids strongly and statically, which we achieved through biotinylation of KaiB (Fig. 1A,B) and the use of streptavidin-coated colloids (Fig 1E-G). We next needed to ensure that biotinylation of KaiB did not intereferre with oscillatory assembly of the KaiBC complex (in the presence of KaiA), and that biotinylated KaiB (b-KaiB) could still bind to KaiC in a phosphorylation-dependent manner. To achieve the former, we used a fluorescence polarization assay to monitor rhythmic complex formation in the KaiABC reaction, finding that the reaction could tolerate the majority of the KaiB proteins being replaced by b-KaiB while still producing high amplitude rhythms (Fig. 1C). Additionally, in pull-down experiments, we found that b-KaiB retains its ability to interact with KaiC as well as its preference for the pS431 state (Fig. S1).

We next aimed to test the ability of KaiC to selectively crosslink b-KaiB-coated colloids via KaiBC complex assembly. To this end, we mixed b-KaiB into a suspension of 1-µm diameter streptavidin-coated colloids, then added either the pS KaiC mutant or the pT KaiC mutant to the suspension. pS and pT are mutated at the phosphorylation sites of KaiC to mimic either a state that permanently allows KaiB binding (pS—S431E;T432A, corresponding to the night phase) or prevents binding (pT—S431A;T432E, corresponding to the morning phase). By imaging the fluorescent-labeled colloids, we found that the colloids remained largely as isolated microspheres in the presence of the non-binding pT mutant, exhibiting no preferential self-association even after a day of incubation (Figs. 2A, S2). This minimal self-association is similar to that observed without b-KaiB (Fig. S2), indicating that non-specific crosslinking is low. In contrast, Fig. 2B shows that b-KaiB-coated colloids incubated with the binding-competent pS KaiC mutant grow into large colloidal aggregates. This crosslinking mechanism is robust, producing qualitatively similar effects with different sized colloids (Fig 2C,D), and forming structures that are system-spanning and relatively immobile compared to pT KaiC samples in all cases (Figs. S2, S3). Thus, b-KaiB can act as a potent material crosslinker that functions only in the presence of appropriately phosphorylated KaiC. To demonstrate that these mesoscopic structural changes can translate to macroscopic material changes visible to the naked eye, we imaged colloidal suspensions undergoing sedimentation in capillaries on the centimeter scale over the course of a day. We observed macroscopic sedimentation dependent on the phosphorylation of KaiC: the larger microscopic clustering seen for pS is mirrored by pronounced sedimentation of the suspension, while the pT colloids remain suspended (Fig. 2E).

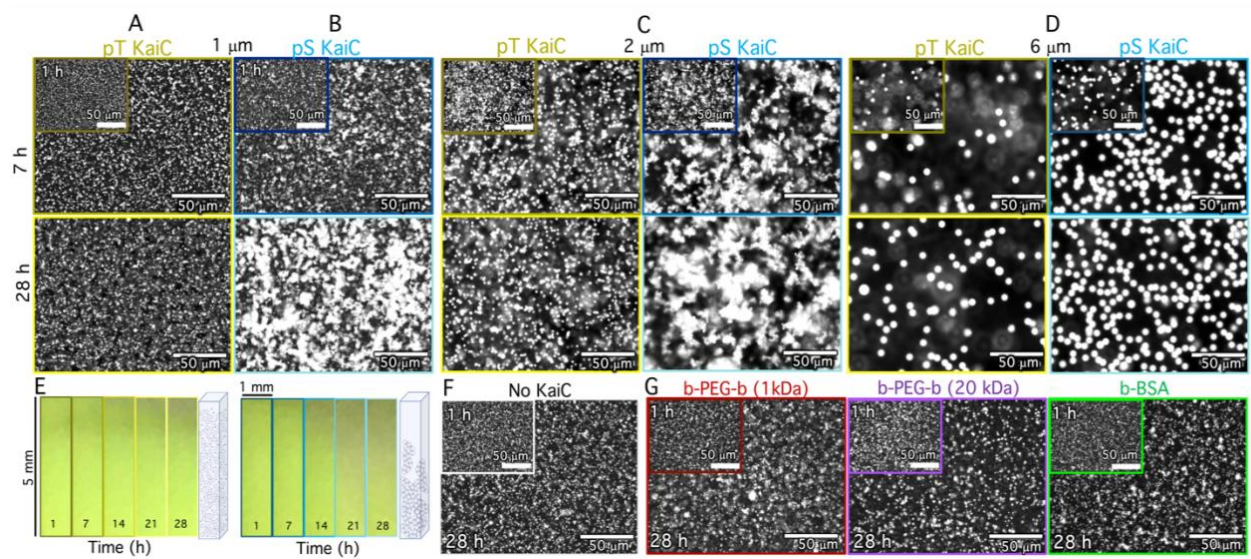

**Figure 2. KaiB-KaiC complexes crosslink colloids with high specificity in a phosphorylation-dependent manner.** (**A,B**) Fluorescence microscopy images of suspensions of 1 µm diameter colloids taken at 1 hr (insets), 7 hrs (top) and 28 hrs (bottom) after mixing with KaiC mutants that are frozen in (**A**) non-binding (pT) or (**B**) binding (pS) states show substantial clustering and assembly of pS-colloids over time that is absent for pT-colloids. (**C,D**) Fluorescence microscopy images of suspensions of colloids of 2 µm (**C**) and 6 µm (**D**) diameter in the presence of pT and pS Kai proteins show that timed aggregation, dependent on the phosphorylation state of KaiC, is preserved for different sizes of colloids. The concentrations of colloids, proteins and reagents, as well as imaging parameters, are identical to those in (A,B). (**E**) Images of a suspension of 1 µm diameter colloids undergoing sedimentation in a capillary (dimensions listed) over 28 hrs in the presence of pT (left) and pS (right) show that pS-colloids sediment more quickly, as indicated by dark regions extending further down the images. The time that each image is captured is listed at the bottom, and the cartoons to the right of panels respectively depict the expected state of the suspension (not drawn to scale). (**F**) The same suspension parameters as in A and B but without KaiC (including only KaiB and b-KaiB) shows minimal clustering over the course of 28 hrs, demonstrating that the KaiB-KaiC complex formation is essential to the colloidal self-assembly shown in A-E. (**G**) Suspensions of streptavidin-coated colloids, with identical conditions to those in A,B, but with Kai proteins replaced with alternative biotinylated constructs that could, in principle, crosslink streptavidin-coated colloids: (left) 1 kDa biotin-PEG-biotin with 1 biotin on each end, (middle) 20 kDa biotin-PEG-biotin with 1 biotin on each end, and (right) biotin-BSA with 8-16 biotins. For all cases the crosslinker molarity matched the KaiC molarity used in A-E, and minimal clustering is observed over 28 hrs, demonstrating that the effect shown in A-F is unique to the KaiB-KaiC binding interaction.

To further demonstrate the robustness and specificity of this self-assembly process, we performed experiments in which we: (1) removed KaiC (Fig 2F), (2) replaced b-KaiB with wild-type KaiB (Fig S2), and (3) replaced streptavidin-coated colloids with passivated non-functionalized colloids (Fig S4). We observed no clustering in any of these control experiments. We also performed experiments in which we replaced Kai proteins with BSA or PEG polymers that have multiple biotinylation sites, but are present on the same linker molecule rather than brought together by macromolecular complex assembly, as in the Kai protein system. While, in principle, these polymer linkers have the potential to act as crosslinkers to bind streptavidin-coated colloids, we found minimal clustering for all linker sizes and number of biotin sites (Fig 2G, S5). Thus, the self-assembled structures we observe specifically require the formation of KaiBC complexes.

***Crosslinking proceeds with kinetics programmed by clock proteins.*** To characterize the clock-driven clustering kinetics that program the colloidal suspensions to transition from freely floating particles to large, connected super-structures, we collected images of the emerging clusters at nine different time-points over the course of a day. To promote mixing and limit colloid settling and sticking, we kept all samples under continuous rotation between imaging intervals. Overlaying temporally color-coded images from these time-course experiments confirms that structure emerges over time in the pS KaiC sample, while minimal clustering is seen in the presence of pT (Fig. 3A-B).

To quantify the time-dependent colloidal self-assembly that is apparent in microscopy images, we use spatial image autocorrelation (SIA) analysis to measure the average size of colloidal clusters at each time point. SIA quantifies the correlation $g(r)$ between pixels separated by a radial distance $r$ (Figs. 3C, S6), which decays exponentially from $g(0) = 1$ with the decay rate indicating the characteristic size of features in an image. Slower decay of $g(r)$ with $r$ indicates larger features (i.e. clusters), as seen for pS compared to pT and long compared to short times (Figs. 3C,S3). By fitting each $g(r)$ curve to an exponential decay, we quantify a characteristic correlation lengthscale $\xi$ of the colloidal system, indicated by the distance $r$ at which the dashed horizontal line intersects each curve in Fig. 2C. We also implemented alternative image analysis algorithms to assess clustering, including quantifying the distribution of pixel intensities (Fig 3D,E, S6) and directly detecting clusters as connected regions in a binarized image (Fig S6), yielding similar results to SIA (Fig. 3D,E, S4). Specifically, the full pixel intensitiy distribution width at 1% (FW1%) and median cluster size both display similar time-dependence as $\xi$ (Fig 3F). To directly compare these different clustering metrics, we normalize each quantity by the corresponding initial value for pT such that the values indicate the degree of clustering, which is one in the absence of clusters (Fig 3F).

Given that pS KaiC is locked into a binding-competent state, the gradual self-assembly of colloids over many hours, suggests that the rate-limiting step in self-assembly is KaiB-KaiC complex formation. Indeed, KaiBC complexes are known to form on the timescale of many hours, likely due to both the slow ATPase cycle in the KaiC CI domain[33] and the time required for KaiB to refold into an alternative fold-switched structure[34,35]. To test this hypothesis, we measured the kinetics of the KaiBC interaction using fluorescence polarization of labeled KaiB, which increases with increasing formation of KaiBC complexes, and compared to the kinetics of material self-assembly. Fig. 3F shows in overlay the time evolution of the relative fluorescence polarization (FP), demonstrating that KaiBC complex formation grows approximately linearly for the first 15 hours after which it approaches saturation, likely reflecting a regime where the majority of both KaiB and KaiC molecules are in complex and have been depleted from solution. The agreement between the kinetics of KaiBC interactions and material self-assembly shown in Fig 3F, as well as the robust specificity of the colloidal assembly (Fig 2A,B,F), is strong evidence that the biochemical properties of the Kai proteins, such as the KaiC catalytic cycle, are regulating the rate of cluster growth.

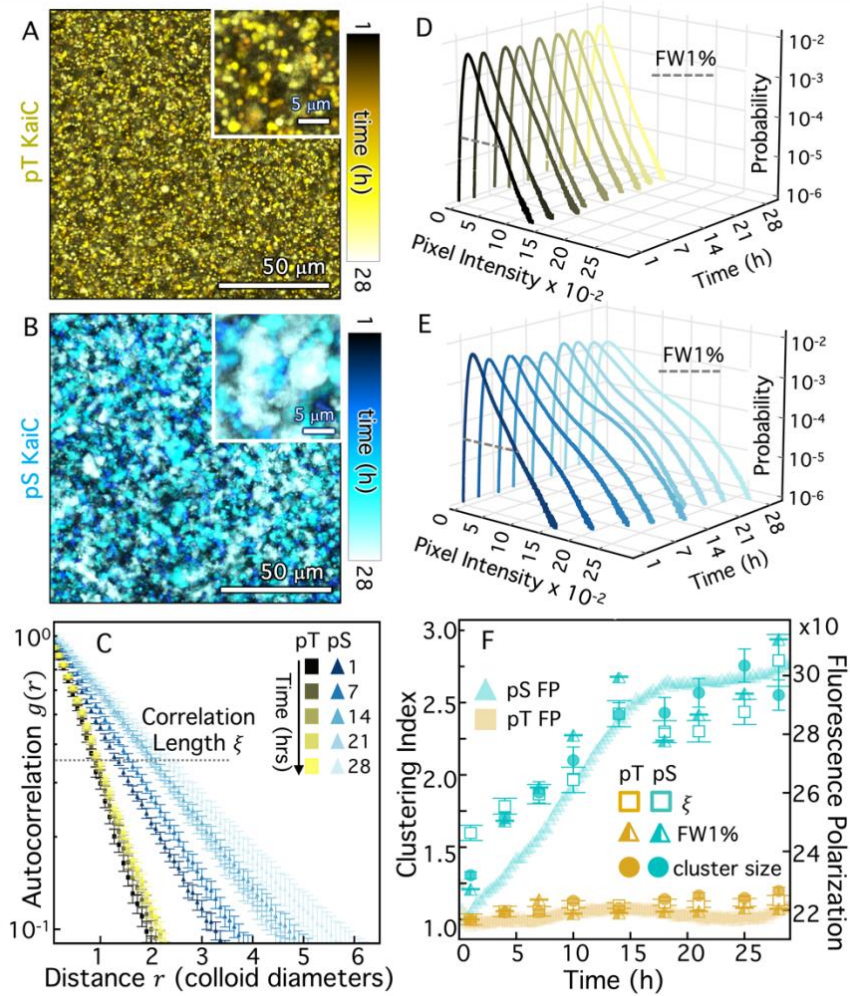

**Figure 3. KaiBC crosslinking mediates robustly timed self-assembly of colloidal clusters that is synced with KaiBC complex formation.** **(A,B)** Colorized temporal projections of time-lapses of pT KaiC (A, yellow) and pS KaiC (B, cyan) over the course of 28 hrs, with colors indicating increasing time from dark to light according to the colorscales. (Insets) Zoomed-in regions of the projections highlighting pS-specific cluster growth over time that is absent for pT. **(C)** Spatial image autocorrelation functions $g(r)$ versus radial distance $r$ (in units of colloid diameter) for 5 different times between 1 and 28 hrs for pT (yellow) and pS (cyan) with color shade indicating time according to the legends in A,B. The characteristic correlation length $\xi$, determined by fitting each $g(r)$ curve to an exponential function is denoted by the intersection of the dashed horizontal line at $g = e^{-1}$. **(D,E)** Pixel intensity probability distributions for pT (C, yellow), and pS (D, cyan) at different times over 28 hrs, with lighter shades denoting later times according to the color scales in A,B. Distributions show broadening and emergence of high-intensity peaks at later times for pS. Dashed grey line denotes the full width at 1% of the maximum probability (FW1%), which serves as a clustering metric used in (F). **(F)** Correlation lengths $\xi$ (open squares), FW1% (half-filled triangles), and median cluster size (filled circles, see Fig S4), each normalized by their initial pT value, show that the time-course of cluster assembly over 28 hrs for pT (gold) and pS (cyan) correlate with the fluorescence polarization (FP) of fluorescently labeled KaiB (right axis, mP), which serves as a proxy for KaiBC complex formation. Both the degree of clustering and FP remain at a minimum for pT, while for pS, both steadily increase for the first ~15 hrs.

***Brownian Dynamics simulations recapitulate timing of cluster formation.*** The correlation of the KaiB fluorescence polarization with the clustering of colloids suggests that Kai protein interaction controls the kinetics of clustering. In order to assess this mechanism, we developed a numerical simulation that captures the key components of our experimental system (see Methods and SI). In the simulations, 1-µm diameter colloids move via Brownian motion in a 50 µm x 50 µm 2D plane, and, when the surfaces of two colloids are within 10 nm of each other (comparable to the size of the KaiBC complex[36]), they can form a bond mediated by b-KaiBC complexes. KaiB and KaiC are assumed to be present at constant concentrations and their interaction to form crosslinks is treated phenomenologically. The probability of complex formation during an encounter is a constant value chosen to match the solution binding kinetics (see SI). We allow simulations to run for 30 hrs and capture the state of the colloids at the same time intervals as in experiments (Fig. 4).

To model our experimental pS KaiC and pT KaiC colloidal systems, we consider cases in in which, respectively, bonds between colloids are incapable of releasing once they are formed (Fig. 4B) and bond formation probability is zero (Fig. 4D). The color-coded temporal overlays of simulation images with 'permanent bonds' (*P*) and 'no bonds' (*N*) show qualitative similarities with the experimental overlays of pS and pT (Fig. 4C,E). To quantitatively compare simulated and experimental cluster assembly kinetics, we perform the same SIA analysis that we use for experimental images (Fig. 3C) to compute time-dependent autocorrelation curves (Fig. 4F) and corresponding correlation lengths (Fig. 4G). Similar to the $g(r)$ trends we observe for experimental pT and pS images (Fig. 3C), Fig. 4F shows that $g(r)$ for the 'no bonds' system exhibits minimal time-dependence and fast decay with distance $r$, indicative of small features that do not change size over time. Conversely, $g(r)$ for 'permanent bonds' (*P*) decays more slowly than *N* at all time-points and broadens substantially over time, indicative of larger clusters that grow over time. The time-course of the corresponding correlation lengths of the 30-hr simulation are likewise similar to the experimental trends in Fig. 3F, with $\xi$ values for the *P* case growing over time and transitioning to slower increase in the latter half of the simulation.

The continued cluster growth for pS (experiments) and *P* (simulations) is somewhat unexpected given the saturation of the fluorescence polarization at ~15 hrs. Specifically, FP saturation suggests that all possible KaiBC complexes have formed, while the colloid data suggest that clusters continue to form and grow after this saturation. To shed light on this seeming paradox, we compute the colloid connectivity number (CCN) from simulations, which measures how many neighboring colloids are connected to a single colloid. Because of the 2D geometry and the size of the colloids, the maximum possible CCN is six. Fig. 4I shows that CCN increases to saturating levels over the course of ~10-15 hrs, similar to the KaiBC FP data, while the simulated correlation lengths continue to increase after this time, albeit less dramatically than the first half of the time-course (Fig. 4G). These data indicate that the rate-limiting step in colloid crosslinking is the assembly of KaiBC complexes rather than the time needed for colloids to come into close contact.

Our results further indicate that cluster growth can proceed even when the majority of colloids are saturated with permanent crosslinks. Such assembly kinetics may arise if the majority of colloids are on the interiors of clusters and saturated, while those on the boundaries may have available b-KaiB binding sites to crosslink to other colloids on the edges of neighboring clusters. Self-assembly thus transitions from that of single colloids coming together to form clusters, to one in which most colloids are participating in clusters that then merge to form larger super-structures. Fig. 4I corroborates this physical picture by comparing the kinetics of cluster formation in the experimental and simulation data with the KaiBC assembly kinetics. The similarity in the shapes of the experimental and simulation curves indicates that the model is indeed capturing the underlying process generating clusters. The clear shift in kinetics at ~15 hrs in all data further corroborates the robustness of the simulations, and demonstrates that self-assembly is rate-limited by the timescale of KaiBC complex formation.

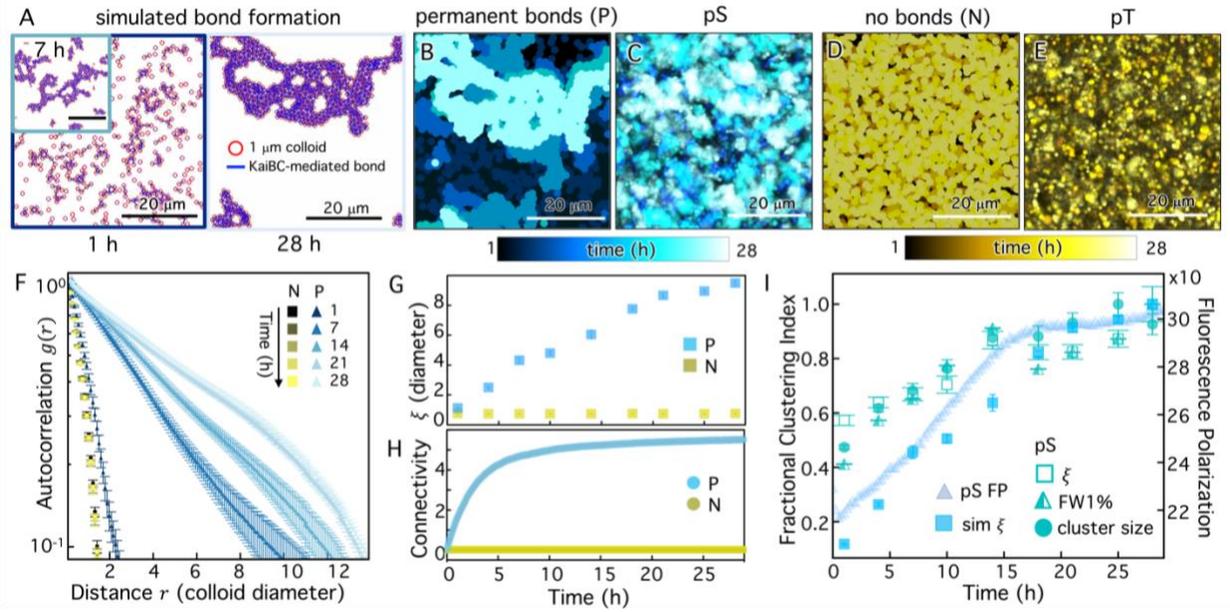

**Figure 4. Kinetic simulations of Kai-mediated crosslinking recapitulate slow formation of colloidal clusters.** (**A**) Simulation snapshots showing clustering of colloids (red circles) crosslinked by permanent bonds (blue lines), analogous to the experimental pS-colloid system, at 1 (left), 7 (inset) and 28 (right) hrs. (**B,C**) Colorized temporal projections of (**B**) simulation snapshots for colloids with permanent crosslinker bonds (permanent bonds, P) and (**C**) experimental snapshots for pS-colloids show similar features emerging over the course of a day. (**D,E**) Colorized temporal projections of (**D**) simulation snapshots for colloids with no crosslinker bonds (no bonds, N) and (**E**) experimental snapshots for pT-colloids both show minimal clustering or restructuring over the course of a day. Times and color-coding used in projections are the same as in Fig. 3, as indicated by the color scales. (**F**) $g(r)$ computed for simulation snapshots, taken at times specified in the legend, for colloids with no bonds (N, yellow squares) and permanent bonds (P, cyan triangles). (**G,H**) Time-course of the (**G**) correlation lengths $\xi$ and (**H**) colloid connectivity number CCN determined from simulations with permanent bonds (P, cyan) and no bonds (N, gold). (**I**) Multiple metrics of clustering and self-assembly resulting from permanent crosslinker bonding in experiments (pS) and simulations (P), each normalized by its maximum value to indicate the fractional clustering index (left axis) measured using each data type. Metrics include: experimental correlation lengths ($\xi$, open squares), simulated correlation lengths (sim $\xi$, filled squares), full width at 1% (FW1%, half-filled triangles), and median cluster size (cluster size, filled circles). Trends in both simulation and experimental data track with the time-course of KaiB fluorescence polarization (right axis (mP), translucent triangles) in a reaction with pS KaiC.

***Oscillations in colloidal clustering depend on the crosslinker density.*** Having demonstrated that material self-assembly can be temporally programmed by the phosphorylation state of the circadian clock proteins, we now investigate the effect of oscillatory interactions between KaiB and KaiC in the wildtype system when KaiA is present. To achieve oscillatory crosslinking, we replace the phosphorylation-locked mutants with WT KaiC and add KaiA, creating a circadian rhythm in both KaiC phosphorylation (mediated by KaiA) and the KaiB-KaiC interaction (Fig. 1C,D).

To guide our experiments, we first aimed to understand how oscillating crosslinkers may translate to the dynamics of material self-assembly. To do so, we extended our model shown in Fig 4 to allow sinusoidally varying colloid binding and unbinding rates (see Methods, SI). In brief, we consider the same binding rate amplitude $p_o$ as in the permanent bond case but we incorporate an oscillation of this rate, $p_{on} = p_o \cos^2(\pi t/T)$, where $T$ is the oscillation period. This construction models the coherent bulk oscillations in the biochemical properties of the KaiABC reaction. We also add a dissociation rate with the same amplitude and functional form as the binding rate, but that is $\pi/2$ radians out of phase, i.e., $p_d = p_o \sin^2(\pi t/T)$. This framework assumes that each connection between two colloids is bonded by a single KaiBC complex ($n = 1$). Fig 5A shows that this minimal bonding allows for oscillatory connectivity, with peaks in CCN observed at times that roughly correlate with the measured peaks in KaiC phosphorylation (Fig 5D). However, the peak CCN values are low compared to the saturating value of 6, and non-zero CCN values are only maintained for a small fraction of the oscillation period, suggesting very weak oscillatory clustering.

However, given the saturating level of Kai proteins in our experiments (~$10^5$ b-KaiB proteins per colloid) and the two orders of magnitude smaller size of the crosslinkers compared to the colloid surface area, we anticipate that more than one KaiBC bond participates in a typical connection between two colloids in experiments. To incorporate multiple bonds per connection into our simulations we modify the dissociation rate to include the number $n$ of bonds that participate in each colloid connection, as $p_d = p_0^n \sin^2(\pi t/T)$. Additional curves in Fig 5A show that $n = 2$ and $n = 3$ produce saturating connections that are unable to appreciably dissociate during a bond oscillation cycle. However, for intermediate cases $n = 4/3$ and $5/3$, we observe pronounced oscillations in connectivity, suggesting similar oscillatory clustering of colloids.

Similar to Fig 4, we translate connectivity to clustering kinetics by computing the correlation length for each time point that is captured in experiments. To compare the time-dependence of complex formation for different bond numbers we evaluate the fractional clustering index, which we define as the baseline-subtracted correlation length, $\xi(t) - \xi_{min}$, normalized by the corresponding maximum value, $\xi_{max} - \xi_{min}$: $FCI = (\xi(t) - \xi_{min})/(\xi_{max} - \xi_{min})$. All values of this function lie between 0 and 1 to allow us to isolate the time-dependence of the clustering. Fig 5B shows that robust oscillatory clustering is achieved for $n = 4/3$ and $5/3$, with the initial peak being more pronounced for $n = 5/3$. Fig 5C shows the simulation snapshots that correlate with the peaks and troughs of the clustering index, visually demonstrating the presence of large super-structures at the peaks and minimal clustering at the troughs.

We understand this complex dependence on bond density as follows: a low density of crosslinkers (i.e., $n = 1$) does not allow superstructures to form, simply because many particles will not be able to find an attachment point, even if the Kai proteins in the system are in a binding-competent state. However, at high crosslinker density (i.e., $n \geq 2$), multivalent effects prevent superstructures from easily disassembling once formed, even when the KaiBC binding probability falls. Thus, the model predicts a 'sweet spot' in crosslinker density (i.e., $1 < n < 2$) where the underlying molecular rhythm in KaiBC interaction will be transduced into material properties with high amplitude (Fig 5B,C).

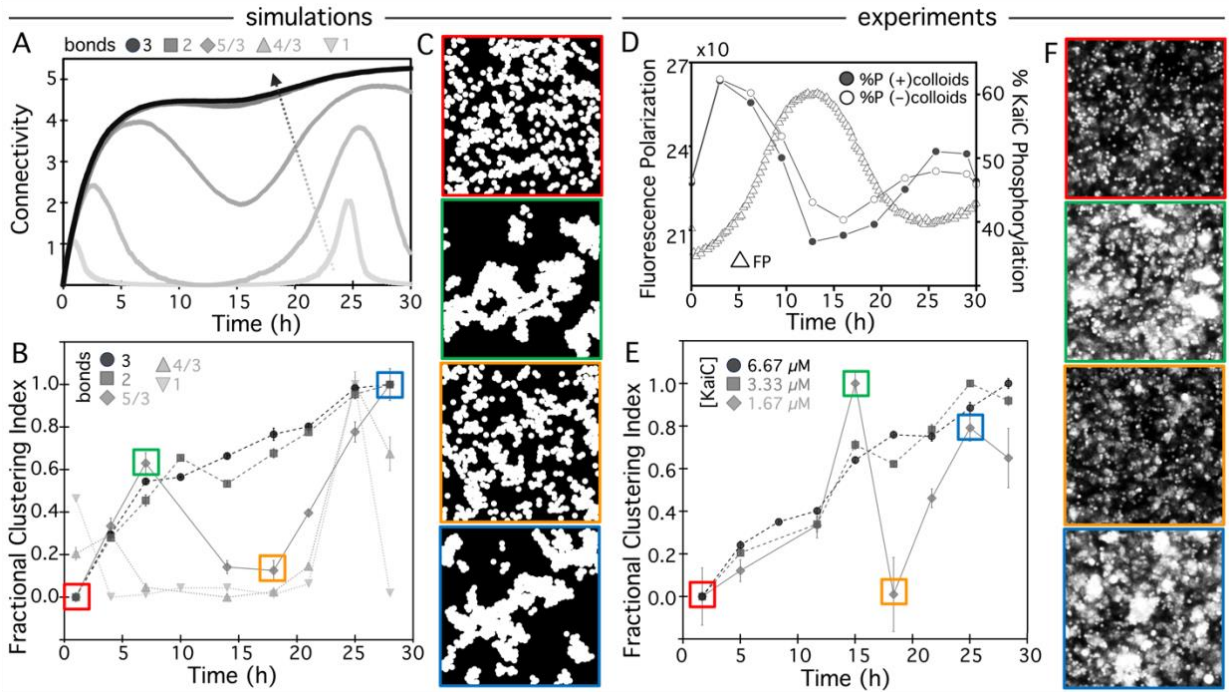

**Figure 5. Circadian oscillation in material properties depends on crosslinker density**. (**A-C**) Simulations model oscillatory colloidal crosslinkers with different numbers of KaiB-KaiC complexes ($n$, bonds) participating in each connection between colloids. (A) Colloid connectivity (CCN) versus time for systems with different numbers of bonds per colloid connection, from light to dark grey: $n = 1, 4/3, 5/3, 2, 3$. Arrow indicates direction of increasing $n$. Intermediate bond numbers $1 < n < 2$ result in oscillating connectivity, while $n = 1$ is not sufficient for pronounced clustering and $n \geq 2$ promotes sustained cluster growth with minimal dissolution. (**B**) The fractional clustering index (see text) versus time for bond densities shown in (A) reveal oscillatory clustering for $n < 2$ that is most pronounced for $n = 5/3$. The colored boxes enclose the data points corresponding to the simulation images with color-matched borders shown in (**C**). (**C**) Simulation snapshots that correspond to troughs (red, orange) and peaks (green, blue) shown in B demonstrate that peaks and troughs correspond to substantial and minimal clustering, respectively. (**D**) Fluorescence polarization (FP) of KaiB (left axis, triangles) and percentage of phosphorylated KaiCs (%P, right axis, circles) during a KaiB-KaiC reaction. %P measurements were performed in the presence (filled circles) and absence (open circles) of colloids, showing that oscillatory KaiC phosphorylation dynamics are unaffected by the presence of colloids. (**E,F**) The fractional clustering index versus time for colloid experiments performed with KaiC concentrations of 6.67 μM (1×, dark grey circles), 3.33 μM (0.5×, grey squares) and 1.67 μM (0.25×, light great diamonds) reveal oscillatory clustering for the lowest concentration, similar to the simulated $n = 5/3$ case, while the two higher concentrations steadily become increasingly clustered over time, similar to the simulated $n \geq 2$ cases. Colored boxes enclose the data points corresponding to the microscope images with color-matched borders shown in (**F**). (**F**) Microscope images that correspond to troughs (red, orange) and peaks (green, blue) shown in E show strong similarities to simulated images and demonstrate minimal and substantial clustering, respectively. All simulated and experimental images shown in (C) and (F) are 50 μm × 50 μm.

Armed with these predictions, we performed experiments at different Kai concentrations, to mimic varying $n$ values in simulations. We first aimed to demonstrate that the ~24 hr oscillation in KaiBC complex formation is not disrupted by the presence of colloids. Fig. 5D confirms that the expected oscillation in KaiBC complex formation, measured by fluorescence polarization, is unperturbed by the buffer conditions used for assembling colloidal materials; and that the oscillatory phosphorylation of KaiC is similarly preserved and largely unaffected by the presence of streptavidin-coated colloids.

We then performed the same full time-course of microscopy measurements as for the pS and pT mutants (Fig 2) at 1×, 0.5×, and 0.25× of the Kai concentration used in these experiments (6.66 µM, Figs 2,3). Evaluating the same fractional clustering index as in simulations (Fig 5B), we find that for the two higher concentrations, colloid superstructures assemble slowly over the course of the day, but show no detectable disassembly, similar to the $n \geq 2$ simulations (Fig. 5B,E). However, at the lower Kai concentration (0.25×, 1.67 µM), oscillatory clustering appears (Fig. 5E,F), with peak and trough times corresponding approximately to the peak and trough KaiBC interaction detected by FP (Fig. 5D). The clustering, dissolution and re-clustering quantified by the *FCI* (Fig 5E) can be observed in the microscope images (Fig 5F) that have remarkably similar features to the simulated images (Fig 5C).

These results demonstrate the achievement of oscillatory assembly and disassembly of a material and the power of predictive modeling to identify the appropriate region of phase space to achieve this engineering feat. Moreover, while oscillatory material assembly indeed represents a transformative advance in materials design, we point out that the robust time constant associated with the material assembly, that is dictated by KaiBC complex formation, provides an additional technological advance. Indeed, the kinetics of cluster assembly are robustly regulated, nearly independent of protein concentration, when we use non-oscillating mutants to program the assembly phase of the material (Fig. S8). This robustness of timing against fluctuating concentration would be difficult to achieve via other assembly control mechanisms (Fig. S9).

**Outlook**

Biomolecular signaling systems typically must maintain their function with high fidelity while interacting with numerous other components crowded within living cells and while subject to unpredictable fluctuations in their environment. These constraints equip networks of interacting biomolecules with unique robustness properties that may allow them to be harnessed to endow synthetic materials and systems with functionality, programmability and autonomous reconfigurability. However, coupling biomolecular systems to synthetic materials to impart desired properties remains a grand challenge in active matter and biomaterials research[13,37–40]. Here, we break new ground by using the KaiABC circadian clock as a prototypical example of a robust biomolecular signaling system. We demonstrate that this system can maintain its natural activity even when functionalized to act as a material crosslinker, and that it can then be used to autonomously regulate timed material self-assembly and oscillation. Specifically, we show that Kai proteins can assemble colloidal suspensions into networks of mesoscopic clusters at rates and efficiencies that are controlled by the phosphorylation state of KaiC. These molecular interactions translate to bulk changes in the sedimentation properties of the materials, visible by the naked eye. Moreover, our mathematical models show that the valency of the clock protein crosslinkers can be used as a switch to allow either sustained self-assembly or rapid dissolution of the material. In intermediate regimes of valency, this system drives robust circadian oscillations in material properties.

This proof-of-concept opens the door to Kai-mediated scheduled crosslinking of a diversity of synthetic and natural materials, such as hydrogels, polymeric fluids, cellulose, and granular materials, to drive user-defined autonomous changes in material properties on a robust programmable schedule. Importantly, the timing of material self-assembly can not only be robustly programmed by Kai clock proteins, but that timing can be precisely tuned with the use of KaiC mutants that operate on cycles of different durations from ~18 hrs –150 hrs[28]. The intrinsic temperature compensation property of KaiC would further ensure that the kinetics of assembly are robust against environmental fluctuations[41]. Other accessory proteins, including SasA and CikA can be incorporated and functionalized to allow for material interactions peaking at other phases of the cycle[16].

These designs can potentially be used to create technologies such as dynamic filtration and sequestration devices, self-healing infrastructure, and programmable wound suturing. Beyond material crosslinking, the Kai system could be used as a synthetic scaffold to gate enzymatic activity to control the release of drugs or achieve metabolic channeling by enforcing spatial proximity between other entities. Beyond time-keeping, biological systems are capable of many information processing tasks including thresholding, fold-change detection, and sign-sensitive filtering of input signals. Because these systems all function based on high molecular specificity, they represent a natural library of computational devices that can be coupled to non-biological systems to achieve autonomous control.

**Methods**

Complete methods and materials are provided in the Supplementary Information. Key details are provided below.

***Protein preparation and characterization:*** KaiA, KaiB, KaiC, pT KaiC (KaiC-AE; S431A, T432E), and pS KaiC (KaiC-EA; S431E, T432A) were recombinantly expressed and purified as previously described[33,42]. Purified proteins were buffer-exchanged into Kai buffer containing 10% glycerol, 150 mM NaCl, 20 mM Tris-HCl (pH 8.0), 5 mM $MgCl_2$, 0.5 mM EDTA (pH 8.0), and 1 mM ATP (pH 8.0). KaiB was functionalized with biotin (b-KaiB) using EZ-Link-Sulfo-NHS-LC-Biotin (ThermoFisher) at a 50× molar excess to KaiB (Fig. 2A-C). The pull-down assay to assess specific binding of KaiC to b-KaiB (see Fig. 2D,E) included 6.5 µM KaiC (wild-type, pT mutant or pS mutant) and 5.5 µM KaiB (55% b-KaiB, 45% unlabeled KaiB) in Kai buffer. Following 8-hr incubation at 30°C, b-KaiB and its binding partners were removed from solution using streptavidin-coated magnetic beads (Cytiva). The resulting supernatant was analyzed by SDS-PAGE (Fig. S1).

To characterize the tolerance of the standard oscillator reaction to b-KaiB, we measured the fluorescence polarization of KaiB in reactions with 1.5 µM KaiA, 3.5 µM KaiC, different ratios of KaiB and b-KaiB, and 0.2 µM FITC-labeled $KaiB^{K25C}$ over the course of 144 hrs, similar to the procedure used previously[43]. We performed the same assay to characterize protein function under colloid-linking conditions which include 5.5 µM KaiB (55% b-KaiB), 6.5 µM KaiC (wild-type, pT or pS), 2.2 µM KaiA, and 0.4 µM FITC-labeled $KaiB^{K25C}$. To characterize the phosphorylation state of KaiC in the colloidal system, KaiC phosphoform composition was resolved by SDS-PAGE analysis. The ratio of phosphorylated KaiC to unphosphorylated KaiC was quantified by gel densitometry.

***Kai-colloid experiments:*** We used 1.0-µm diameter streptavidin-coated polystyrene microspheres (Fluoresbrite YG Polysciences) as the colloids in all experiments. Colloids were washed and resuspended in Kai buffer such that the final concentration in all experiments is 1.26% solids (~$4.0 \times 10^{-5}$ µM). To prepare Kai-colloid suspensions, we add 3.6 µM b-KaiB, 2.9 µM wild-type KaiB, 2.2 µM KaiA, and 6.5 µM of either wild-type KaiC, pS KaiC or pT KaiC. The time of KaiC addition sets $t = 0$ for each experiment. For microscopy experiments, Kai-colloid suspensions were flowed via capillary action into passivated sample chambers consisting of a glass microscope slide and No. 1 coverslip fused together with heated ~120-μm thick parafilm spacers to accommodate 8 μL of sample. After sealing the chambers with UV glue, they were immediately placed on a 360° rotator at 30°C for the duration of each ~30 hr experiment except when being imaged. For each experiment two replicates were prepared and imaged immediately after one another.

Colloidal suspensions were imaged using an Olympus IX73 epifluorescence microscope with a 40× 0.6 NA objective, 480/535-nm excitation/emission filters, and a Hamamatsu ORCA-Flash 2.8 CMOS camera. For each condition and time-point, 6 images were captured in equidistant regions in the sample chamber within a span of 5 mins. All data shown consists of 2-3 replicate experiments with vertical error bars representing standard error. To construct confocal z-stacks (Fig S2), we used a Nikon A1R scanning confocal microscope with a 60× 1.4 NA oil-immersion objective, a 488 nm laser and 488/595 nm excitation/emission filters.

We processed and analyzed images, as depicted in Fig. S6, using custom-written Python codes[44–46]. We evaluated the distribution of pixel intensities across all images for a given time and condition, which we normalize to probability density distributions. We performed spatial image autocorrelation (SIA) analysis in Fourier space and directly measured 2D cluster sizes in real space. For both analyses, we first binarized images using local thresholding algorithms[47]. To measure cluster sizes, we identified each connected set of pixels above threshold as a cluster, counted the number of pixels in each such region, and divided by the cross-sectional area of a colloid. To quantify the distribution of cluster sizes we evaluated the cumulative distribution function (CDF) of cluster sizes. We used the same binarized images to perform SIA, which measures the correlation in intensity values $g(r)$ of each pair of pixels in a given image that are separated by a radial distance $r$[48], and averages over all pairs with a given $r$. Data shown in Fig. 3 are the average and standard error of $g(r)$ curves measured across all images at a given time and condition. We fit each $g(r)$ to an exponential function to quantify a characteristic correlation lengthscale $\xi$ associated with the features (e.g., colloids, clusters) in a given image, which we normalize by the colloid diameter to quantify $\xi$ in terms of the number of colloids it spans.

We performed sedimentation experiments in borosilicate glass capillaries with 1 mm × 1 mm inner cross-section (Wale Apparatus) that accommodate ~10 μL of sample. Colloidal suspensions were pipetted into the capillaries which were then sealed by adhering glass coverslips to the openings using UV-curable adhesive. The capillaries were mounted vertically, illuminated with a white light LED, and imaged every hour for 36 hours using an iPhone 6s.

*Mathematical Modeling and Simulations:* We simulate the dynamics of the experimental system using Brownian Dynamics implemented in C++[49], as described in SI. Our system consists of 500 colloidal particles of diameter $\sigma = 1$ μm, confined to a two-dimensional box with edge length 50 μm and periodic boundary conditions. The colloids occupy a static area fraction of 16%, set to match experimental conditions by evaluating the fraction of pixels above threshold in binarized experimental images. At the beginning of each simulation, all colloids are separate particles undergoing Brownian diffusion in 2D. When the surfaces of two colloids come within a distance $l = 10$ nm of each other (the approximate size of a KaiBC complex[36]), they have a non-zero probability of linking together. We simulate three cases that correspond to our experimental studies: (1) Permanent crosslinking, where, once formed, bonds between colloidal particles are permanent; (2) No crosslinking, where bonds never form between colloidal particles regardless of their proximity; and (3) Oscillatory crosslinking, where bond formation and dissolution follow the oscillatory complexing of KaiB and KaiC. SI Table S1 provides all simulation parameters, their relation to experimental values, and rationale for their choice.

For cases (1) and (3), when a pair of particles are within a center-to-center distance of $r_0 = \sigma + l$, they can become crosslinked with a certain probability. This attachment probability at simulation time $t$ is $p_a = p_0 \cos^2(\pi t/T)$, where $T = 24$ hrs represents the crosslinker oscillation period. The probability amplitude $p_0$ is a phenomenological parameter determined from the fluorescence polarization data for pS KaiC (see

Fig. 3F). In cases where the particles can unlink, we implement a detachment probability, $p_d = p_0^n \sin^2(\pi t/T)$, where $n$ is the number of bonds (KaiABC crosslinkers) connecting the particle pair under consideration. At the beginning of the simulation, the system has a maximum probability of attachment and a minimum detachment probability to replicate the experimental schedule of the KaiB-KaiC interaction. We run simulations for $69120\tau$, corresponding to 48 hrs of experimental time (see SI Movie S1, Table S1), and show averages over 5 runs in the results presented in the paper.

## Acknowledgments

We thank Jeffrey Wang and Eliana Petreikis for their work on the preliminary design of the system, and Katarina Matic, Maya Hendija and Juexin Marfai for assistance with control experiments. We thank Megan Valentine, Ryan McGorty, and Jonathan Michel for insightful discussions. This work was funded by a WM Keck Foundation Research grant and NSF DMREF grant awarded to RMRA, JLR, MJR and MD, and NIH R01 GM107369 to MJR.

## Data Availability

All data will be made freely available upon request.

## Competing Interests

The authors declare no competing interests.

## Author Contributions

Conceptualization: JLR, MD, MJR, RMRA
Methodology: JLR, MD, MJR, RMRA, GL
Investigation: GL, LM, MC, MN, LB, JK, SJK, SB, LF, MJR, MD, JLR, RMRA
Visualization: GL, MC, LM, MN, RMRA
Supervision: JLR, MD, MJR, RMRA
Writing—original draft: GL, LM, MJR, MD, RMRA
Writing—review & editing: GL, MC, JLR, MD, MJR, RMRA

# Timed material self-assembly controlled by circadian clock proteins

*Gregor Leech[1], Lauren Melcher[2†], Michelle Chiu[3†], Maya Nugent[1], Lily Burton[4], Janet Kang[5], Soo Ji Kim[4], Sourav Roy[6], Leila Farhadi[6], Jennifer L. Ross[6], Moumita Das[2,7], Michael J. Rust[5], Rae M. Robertson-Anderson[1*]*

[1]Department of Physics and Biophysics, University of San Diego, San Diego, California 92110, United States

[2]School of Mathematical Sciences , Rochester Institute of Technology, Rochester, New York 14623, United States

[3]Graduate Program in Biophysical Sciences, University of Chicago, Chicago, Illinois 60637, United States

[4]Department of Biochemistry and Molecular Biophysics, University of Chicago, Chicago, Illinois 60637, United States

[5]Department of Molecular Genetics and Cell Biology and Department of Physics, University of Chicago, Chicago, Illinois 60637, United States

[6]Department of Physics, Syracuse University, Syracuse, New York 13244, United States

[7]School of Physics and Astronomy, Rochester Institute of Technology, Rochester, New York 14623, United States

[†]These authors contributed equally to this work.

## Supplementary Information

**Section S1. Supplementary Methods**

**Figure S1. Biotinylation of KaiB preserves phosphorylation-dependent binding to KaiC**.

**Figure S2. Confocal microscopy of colloidal suspensions incubated with pT KaiC + 55% b-KaiB (left), pS KaiC + 55% b-KaiB (right), and pS KaiC + 0% b-KaiB (middle).**

**Figure S3. Complementary image analysis methods quantify selective colloid clustering with pS KaiC**.

**Figure S4. Time-dependence of KaiC-mediated cluster formation quantified via cluster size analysis.**

**Figure S5. Clock-mediated crosslinking of 2 μm and 6 μm colloids display the same self-assembly kinetics as 1 μm colloids in the presence of biotinylated Kai proteins.**

**Figure S6. Brightfield microscopy of suspensions of passivated colloids incubated with pT KaiC (left), pS KaiC (middle), or WT KaiC (right) with 55% b-KaiB.**

**Figure S7.** Streptavidin-coated colloids in the presence of biotinylated PEG or BSA (no Kai proteins), or in the presence of biotin-KaiB and WT KaiB but no KaiC, show no apparent crosslinking or clustering.

**Figure S8.** The time constant for colloidal crosslinking is robust to the concentration and mirrors the time constant of KaiBC complex formation.

**Figure S9.** Both timing and extent of salt-mediated clustering of colloids depend on salt concentration.

**Movie S1.** Animations of simulations of colloids crosslinking under the action of (A) no crosslinkers (N), (B) permanent crosslinkers (P), and in cases in which each connection between two colloids is formed by (C) 1, (D) 2 or (E) 3 oscillatory crosslinkers.

**Table S1.** Table of parameters used in modeling and simulations.

**Supplementary References**

## Section S1. Supplementary Methods

*Protein preparation and characterization*

*Protein expression and purification:* KaiA, KaiB, KaiC, pT KaiC (KaiC-AE; S431A, T432E), and pS KaiC (KaiC-EA; S431E, T432A) were recombinantly expressed and purified as previously described [1,2]. For experiments other than the initial characterization by fluorescence polarization in Fig 2C, WT KaiC carried an N-terminal FLAG epitope and was expressed using a SUMO tag following[3]. Purified proteins were buffer-exchanged into Kai buffer containing: 10% glycerol, 150 mM NaCl, 20 mM Tris-HCl (pH 8.0), 5 mM $MgCl_2$, 0.5 mM EDTA (pH 8.0), 1 mM ATP (pH 8.0). Protein concentration was measured by Bradford Assay (Bio-Rad) using bovine serum albumin (BSA) as a standard.

*Biotinylation of KaiB:* KaiB stock (70 µM) was buffer-exchanged into labeling buffer (pH 8.0), containing 20 mM HEPES and 150 mM NaCl, using Zeba Spin Desalting Columns with 7KD MWCO (ThermoFisher). KaiB was functionalized with biotin using EZ-Link-Sulfo-NHS-LC-Biotin (ThermoFisher). The biotin reagent was added in 50× molar excess to KaiB, and the reaction was incubated at room temperature for 30 minutes. Unincorporated biotin was then removed using Zeba spin columns equilibrated in Kai buffer.

*Pull-down assay of KaiBC complexes using b-KaiB as bait:* Reactions with 6.5 µM KaiC (wild-type, pT mutant or pS mutant) were mixed with 5.5 µM KaiB (55% b-KaiB, 45% unlabeled KaiB) in Kai buffer and incubated for 8 hrs at 30°C to allow protein complex formation. The negative control without b-KaiB (6.0 µM unlabeled KaiB) was incubated for 20 hours. b-KaiB and its binding partners were removed from solution by incubation with streptavidin-coated magnetic beads (Cytiva). To prepare the bead slurry, 20 µL of the resin stock was washed twice with 100 µL of Kai buffer. The resulting pellet was resuspended with with 5 µL of the Kai protein reaction. The reaction was incubated for 30 minutes at 22°C with shaking at 950 rpm, after which the beads were magnetically pelleted. The resulting supernatant was transferred into a new tube and analyzed by SDS-PAGE using 4-20% Criterion TGX gels (Bio-Rad) and stained using SYPRO Ruby (Invitrogen).

*Fluorescence polarization:* To characterize the tolerance of the standard oscillator reaction to b-KaiB, clock reactions were prepared by mixing together 1.5 µM KaiA, 3.5 µM KaiC, different ratios of KaiB and b-KaiB, and 0.2 µM FITC-labeled $KaiB^{K25C}$ in Kai buffer, similar to the procedure used previously[4]. For each reaction, the total concentration of KaiB and b-KaiB was 3.5 µM. 30 µL of the reaction mixture was added to a black 384-well plate, which was sealed with low-autofluorescence polyolefin film (USA Scientific). The plate was loaded into a plate reader to measure fluorescence polarization with 485 nm excitation (20 nm bandpass) and 535 nm emission (25 nm bandpass) using a 510-nm dichroic filter. Measurements were taken every 15 minutes. Wells containing only Kai storage buffer were used as blanks. We performed the same fluorescence polarization assay to characterize protein function under colloid-linking conditions. For these experiments, we mixed together 5.5 µM KaiB (55% b-KaiB, 45% unlabeled KaiB), 6.5 µM KaiC (wild-type, pT, or pS), 2.2 µM KaiA, and 0.4 µM FITC-labeled $KaiB^{K25C}$ in Kai buffer.

*Characterization of phosphorylation state of KaiC in the colloidal system:* KaiC phosphorylation in each sample was resolved by SDS-PAGE analysis using 10% Criterion Tris-HCl gels (Bio-Rad) run for 3 hrs at 125 V. The gels were stained with SYPRO Ruby (Invitrogen) and imaged using ChemiDoc (Bio-Rad). The ratio of phosphorylated KaiC to unphosphorylated KaiC was quantified by gel densitometry.

*Clock-colloid experiments*

*Colloidal suspensions:* We used streptavidin-coated Fluoresbrite YG 1.0-µm diameter polystyrene microspheres (Polysciences) as the colloids in all experiments. The colloids were stored suspended in water at 1.4% solids at 4°C. Immediately prior to each experiment, the colloids were washed and resuspended to 3.7% solids in Kai reaction buffer as follows. 58 µL of the stock suspension was centrifuged in a low-retention microcentrifuge tube at 11500× $g$ for 5 mins to pellet the colloids. The supernatant was immediately removed and replaced with 80 µL of Kai buffer, followed by vortexing to resuspend the colloids. This washing process is repeated two more times. Following the third spin down, the colloids are resuspended at 3.7% solids in Kai buffer. The suspension was mixed by vortexing and then further homogenized by sonication for 45 mins at 4˚C.

To prepare Kai-colloid suspensions, we add 3.6 µM b-KaiB to the sonicated colloidal suspension, vortex for ~3 secs, and incubate at room temperature for 5 mins to allow b-KaiB to coat the colloids. Next, we mix in 2.9 µM unlabeled KaiB and 2.2 µM KaiA (if using) to create a master mix that we then divided into three tubes to which we add 6.5 µM of either wild-type KaiC, pS KaiC or pT KaiC. $t = 0$ is defined as the time of KaiC addition. The final colloid concentration is 1.26% solids (~$4.0×10^{-5}$ µM) in all cases.

*Microscopy samples:* Kai-colloid suspensions are flowed via capillary action into passivated sample chambers which are prepared as follows. Glass microscope slides and No. 1 coverslips are cleaned with methanol, rinsed with deionized water, and air dried. Three 22 × 3 mm flow channels are formed by fusing the slide and slip together with heated ~120-µm thick parafilm spacers to accommodate 8 µL of sample in each channel. To prevent non-specific binding of colloids and proteins to the chamber walls, we passivate the surfaces using bovine serum albumin (BSA)[5] as follows. We fill sample chambers with 10 mg/ml BSA in Kai buffer, incubate in a hydrated chamber at room temperature for 30 minutes, then flush the BSA solution out with fresh Kai buffer. Sample chambers are then filled with the colloidal suspension, sealed with UV-curable adhesive, and placed on a 360˚ rotator at 30 ˚C. For each experiment two replicates of each case are prepared and imaged immediately after one another.

*Microscopy Experiments:* Colloidal suspensions were imaged using an Olympus IX73 epifluorescence microscope with a 40× 0.6 NA objective, 480/535-nm excitation/emission filters, and a Hamamatsu ORCA-Flash 2.8 CMOS camera. 1920×1440 pixel images of suspensions are captured 2-4 µm above the coverslip, in 6 equidistant regions around the horizontal center of sample chambers. Single images were captured sequentially for each sample in the three-lane chamber within a span of five minutes, for a total of 18 images in a 15 min period. This process was then immediately repeated for the replicate chamber. This set of 18 images were captured every ~3 hrs over a 28 hr period. All data shown consists of 2-3 replicate experiments with vertical error bars representing standard error. Horizontal error bars indicate the variation in image acquisition times across replicates binned together into one data point (~15 mins for most cases).

To construct confocal $z$-stacks shown in Figs 3 and 6, we imaged samples using a Nikon A1R scanning confocal microscope with a 60× 1.4 NA oil-immersion objective. Colloids are imaged using a 488 nm laser with 488/595 nm excitation/emission filters. Stacks of 51 square images, with planar edge lengths of 512 pixels (210 µm), are constructed using a 0.2 µm $z$-step size, starting from the coverslip ($z = 0$) and going up to $z = 10$ µm.

*Image post-processing and analysis:* We analyzed images using three different approaches (see Fig 3). All image post-processing and analyses were performed using custom-written Python codes[6–8]. We first evaluated the distribution of pixel intensities across all images for a given time-point and condition, which we normalize to probability density distributions. To account for variations in image brightness across different images we subtract from each pixel value the [mean – minimum] pixel value. We construct histograms using 5000 bins.

We also performed spatial image autocorrelation (SIA) analysis in Fourier space and directly measured 2D cluster sizes in real space. For both analyses, we first binarized images by using local thresholding algorithms with a local block of 1051×1051 pixels[9,10], such that all pixel values are converted to 1 or 0. To measure cluster sizes, we identify each connected set of pixels above threshold as a cluster and count the number of pixels in each such region. By dividing the number of pixels within each 2D area by the maximum cross-sectional area of a colloid, we estimate the size of each cluster in terms of the minimum number of colloids it comprises. To quantify the distribution of cluster sizes we evaluate the cumulative distribution function (CDF) of cluster sizes across all images for a given time-point and condition.

We used the same binarized images to perform SIA, which measures the correlation in intensity values $g(r)$ of each pair of pixels in a given image that are separated by a radial distance $r$ [11], and averages over all pairs with a given $r$. In practice, we generate $g(r)$ for each image by taking the fast Fourier transform of the image $F(I(r))$, multiplying by its complex conjugate, applying an inverse Fourier transform $F^{-1}$ and normalizing by the squared intensity $[I(r)]^2$: $g(r) = [F^{-1}|F(I(r))|^2]/[I(r)]^2$. We generate a separate $g(r)$ curve for each image, and the data shown in Figs 3,4 and 6 are the average and standard error of $g(r)$ across all images at a given time and condition. To quantify a characteristic lengthscale associated with the features (e.g., colloids, clusters) in a given image, we fit the corresponding $g(r)$ to an exponential function, $g(r) = (1-A)e^{-r/\xi} + A$, where $A$ is a constant that accounts for non-zero asymptotes and $\xi$ is the correlation length that estimates the average size of clusters in an image. To generate the experimental data shown in Figs 4-6, we compute $\xi$ for each image, average over all images for a given condition and time-point, and normalize by the colloid diameter to quantify $\xi$ in terms of the number of colloids it spans.

*Sedimentation Experiments:* Sedimentation experiments were carried out in borosilicate glass capillaries with 1 mm × 1 mm inner cross-section, 0.2 mm wall thickness and 11 mm length (Wale Apparatus, #8100-050), accommodating ~10 µL of aqueous sample. Capillaries are secured to a 10 mm × 25 mm microscope slide with UV-curable adhesive, oriented such that each end extends ~0.5 mm beyond the slide. The overhanging ends are then sanded so they are flush with the slide. The colloidal suspensions, prepared as described above, are pipetted into the capillaries via capillary action. To seal the capillaries, glass coverslips (7 mm × 22 mm) are adhered to the bottom and top openings using UV-curable adhesive. A second coat of UV-curable adhesiveis added to the junctions between the chambers and coverslip to prevent sample evaporation and leakage.

The samples are mounted vertically against a white background in an enclosed space protected from light. To acquire time-lapse images of the colloidal suspensions undergoing during sedimentation, the mounted capillaries are illuminated with a white light LED and 1024 x 1024 square-pixel images are captured every hour for 36 hours using an iPhone 6s. The capillaries and lighting remain unperturbed for the entire 36-hr acquisition.

*Mathematical Modeling and Simulations*

*Simulation Setup:* We simulate the dynamics of the experimental system using Brownian Dynamics implemented in C++[12,13]. Our system consists of 500 colloidal particles of diameter $\sigma = 1$ µm, confined to a two-dimensional box with edge length 50 µm and periodic boundary conditions. The colloids occupy a static area fraction of 16%, set to match the bead density in the imaging plane estimated from the binarized experimental images at $t = 0$. At the beginning of each simulation, all colloids are separate particles undergoing Brownian diffusion in 2D. When the surfaces of two colloids come within a distance $l = 10$ nm of each other[14], they have a non-zero probability of linking together. If a pair of particles gets too close, they experience hard sphere repulsion. We simulate three cases that correspond to our experimental studies: (1) Permanent crosslinking, where, once formed, bonds between colloidal particles are permanent; (2) No crosslinking, where bonds never form between colloidal particles regardless of their proximity; and (3) Oscillatory crosslinking, where bond formation and dissolution follow the oscillatory activation and deactivation of the KaiBC complex. SI Table S1 provides all simulation parameters, their relation to experimental values, and rational for their choice.

*Crosslinking Kinetics:* For the cases (1) and (3), when a pair of particles are within a center-to-center distance of $r_0$, they can become crosslinked with a certain probability. This probability of attachment at simulation time $t$ is $p_a = p_0 \cos^2(\pi t/T)$, where $T$ represents the crosslinker oscillation period, and is set to 24 hours commensurate with the ~24-hr rhythm of the Kai clock. The probability amplitude $p_0$ is a phenomenological parameter, and we determine its value from the fluorescence polarization data for pS KaiC (see Fig 2). We fit an exponential function $y(t) = y_{ss} + (y_0 - y_{ss})e^{-t/\tau_F}$ to this data, where $y_0$ and $y_{ss}$ denote the fluorescence polarization value at the onset of the experiment and at time when the system has reached a steady state, respectively. We measure $\tau_F \simeq 4 \times 10^4$ s from this fit, which we use to obtain the corresponding probability $p_0 \simeq 1 - e^{-\tau/\tau_F} \approx \tau/\tau_F$. Here, $\tau = 2.4$ s is the timescale used to non-dimensionalize simulations and is set to the time for a colloid to diffuse across its own diameter in water. In cases where the particles can rhythmically unlink, we implement a detachment probability, $p_d = p_0^n \sin^2(\pi t/T)$, where $n$ is the number of bonds (KaiABC crosslinkers) connecting the particle pair under consideration. At the beginning of the simulation, the system has a maximum probability of attachment and a minimum detachment probability to simulate the KaiABC reaction beginning in a phase corresponding to KaiB-KaiC complex formation. We vary $n$ as a proxy for varying Kai concentration.

*Dynamics of colloidal particles:* The movement of the $i^{\text{th}}$ colloidal particle follows the overdamped Langevin equation, $\frac{d\mathbf{r}_i}{dt} = \frac{D}{k_B T}\sum_j(\mathbf{F}_{c,ij} + \mathbf{F}_{LJ,ij}) + \sqrt{2D}\boldsymbol{\eta}_i$, where $k_B$ is the Boltzmann constant, temperature $T$ is set to room temperature, $D$ is the colloid diffusion coefficient, and $\eta$ represents Gaussian noise with zero mean and unit variance. The interparticle interaction force consists of two contributions. The first contribution comes from an elastic spring force $\mathbf{F}_{c,ij}$ due to stretching or compressing the crosslinker that connects particles $i$ and $j$, defined as $\mathbf{F}_{c,ij} = -K(r_{ij} - r_0)\hat{r}$, where $K$ is the spring constant, $r_{ij}$ is the distance between the centers of particles $i$ and $j$, and $r_0$ is the sum of particle diameter $\sigma$ and crosslinker rest length. The second contribution comes from the hard-sphere interaction between particles, modeled by the repulsive part of the 12-6 Lennard-Jones potential, $V_{LJ} = 4\epsilon\left[\left(\frac{\sigma}{r_{ij}}\right)^{12} - \left(\frac{\sigma}{r_{ij}}\right)^6\right]$, where $\epsilon$ is the interaction strength. We non-dimensionalize equations by scaling distances by the particle diameter $\sigma$, times by the diffusion timescale $\tau = \sigma^2/D$, and energies by $k_B T$. We run simulations for $69120\tau$, corresponding to 48 hrs of experimental time (see Table S1), and show averages over 5 runs in the results presented in the paper.

*Spatial image autocorrelation (SIA):* We performed the same SIA algorithms used to analyze experimental images on system configuration images from simulations generated using MATLAB at time points corresponding to the experimental time points of 1, 7, 14, 21 and 28 hrs. Prior to SIA analysis, image files are converted to binary images with colloids being filled white circles and the empty space being black, similar to experimental binarized images. To determine the average $g(r)$ and $\xi$ for each time point, we analyzed images for 5 consecutive time points centered on the time of interest, consistent with experiments. $g(r)$ and $\xi$ values shown in Figs 5 and 6 are averages of the values determined from the 5 different images with error bars representing standard error.

*Simulation animations:* Animations of the simulations for all 5 crosslinking cases considered in the study, provided in SI, were rendered using an OpenGL (Open Graphics Library) script written in MATLAB. The packages used to run them were GCC (C/C++) and ffmpeg, and the inputs were the colloid center positions and colloid IDs that indicated which colloids were crosslinked.

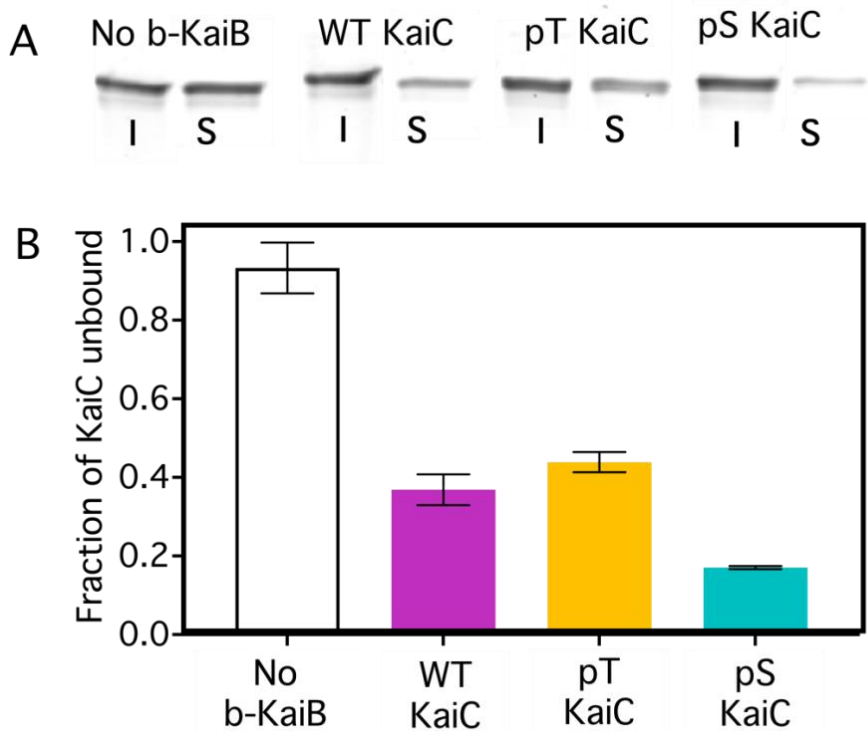

**Figure S1. Biotinylation of KaiB preserves phosphorylation-dependent binding to KaiC**. Streptavidin pull-down analysis of the KaiB-KaiC interaction with and without biotinylated KaiB (b-KaiB). Data from left to right are for: wild-type KaiC with 100% wild-type KaiB (No b-KaiB, white), wild-type KaiC with 55% b-KaiB and 45% KaiB (WT KaiC, magenta), non-binding pT KaiC with 55% b-KaiB and 45% KaiB (pT KaiC, yellow), and constitutively-binding pS KaiC with 55% b-KaiB and 45% KaiB (pS KaiC, cyan). **(A)** SDS-PAGE gel of KaiC remaining in the supernatant (S), and thus not in complex with b-KaiB, following pull-down using magnetic streptavidin-coated beads. Each pair of bands corresponds to KaiC in the input reaction after 8-hr incubation (I) and supernatant following pull-down (S). **(B)** Ratios of the supernatant and input determined by gel densitometry for each pair shown in (A), with 0 indicating that all KaiC is bound to b-KaiB while 1 indicates no KaiC is bound.

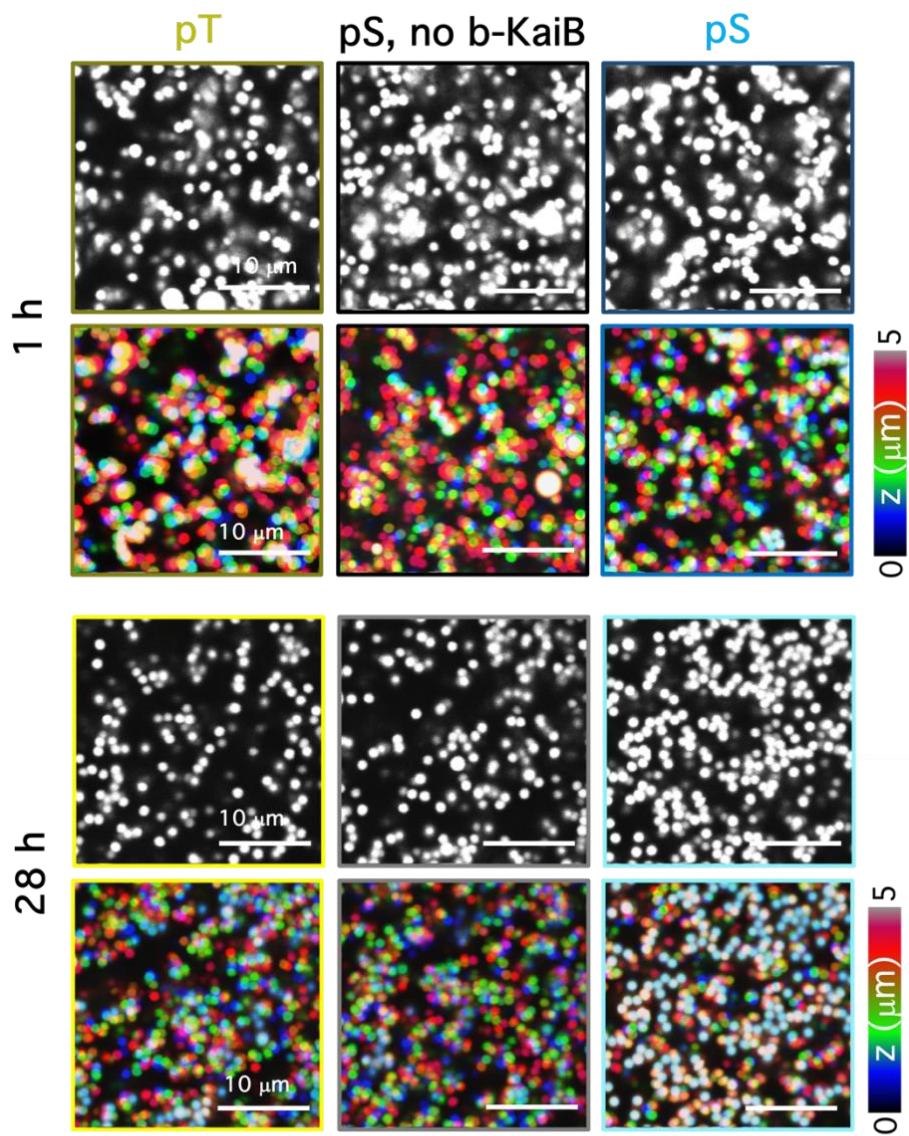

**Figure S2. Confocal microscopy of colloidal suspensions incubated with pT KaiC + 55% b-KaiB (left), pS KaiC + 55% b-KaiB (right), and pS KaiC + 0% b-KaiB (middle).** Confocal images (rows 1,3) and z-projections (rows 2,4) of colloidal suspensions incubated with 6.5 µM mutant KaiC (pT or pS) and 6.5 µM KaiB for 2 hrs (rows 1,2) and 28 hrs (rows 3,4). The left and right columns are with pT KaiC and pS KaiC, respectively, and 55% of the KaiB is b-KaiB (as in the experiments in the main text). The middle column includes the constitutively binding pS KaiC mutant but none of the KaiB is biotinylated. The left and middle columns are nearly indistinguishable from one another and show minimal clustering, indicating that the crosslinking is specifically mediated by b-KaiB-KaiC interactions. The right column shows more clustering of colloids and hindered motion, seen as more white regions in the z-projections. Data acquisition and imaging parameters are described in Methods and SI Methods.

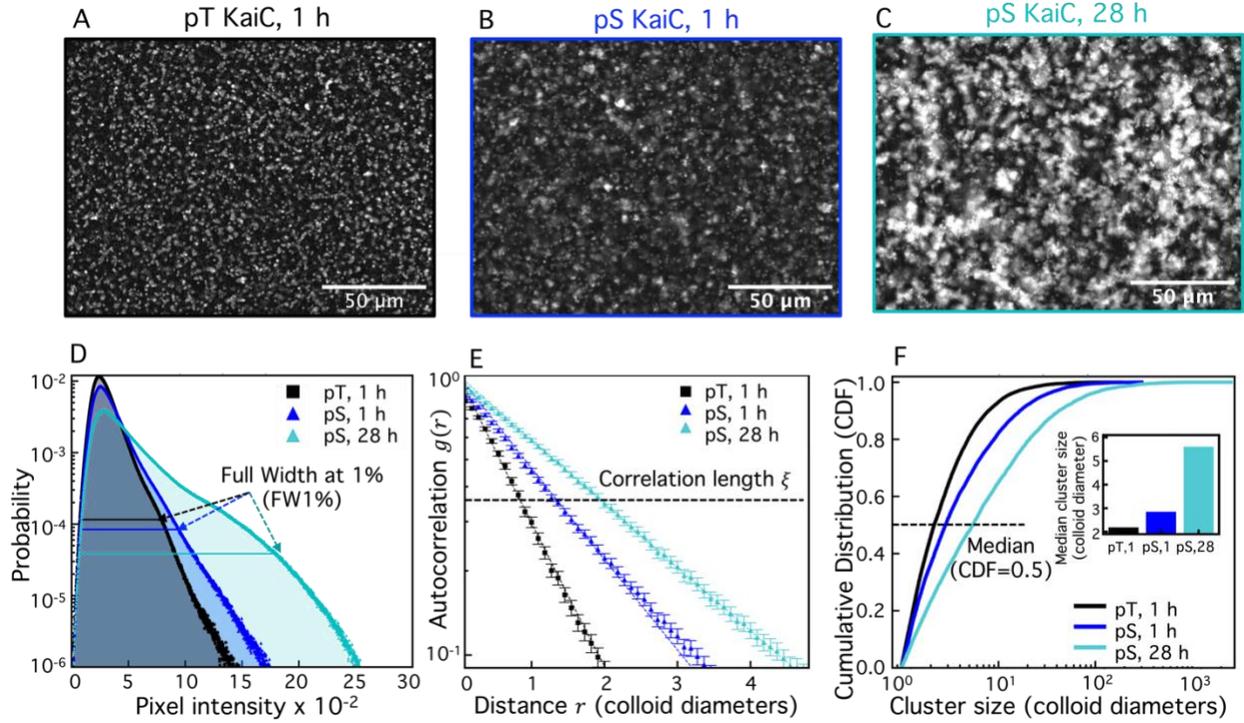

**Figure S3. Complementary image analysis methods quantify selective colloid clustering with pS KaiC**. **(A-C)** Fluorescence microscopy images of 1-µm colloids taken at 1 h (A,B) and 28 h (C) after mixing with KaiC mutants that are frozen in non-binding (A, pT) or binding (pS, B,C) states. **(D-F)** Quantification of colloidal clustering in epifluorescence images by evaluating **(D)** the pixel intensity distributions of images, **(E)** spatial image autocorrelation (SIA) function $g(r)$, i.e., the correlation between two pixels separated by a radial distance $r$, and **(F)** the normalized cumulative distribution (CDF) of cluster sizes, where a cluster is defined as a connected set of above-threshold pixels. To compare these complementary analyses, we compute **(D)** the full width of each intensity distribution at 1% of the corresponding mode, denoted by the horizontal lines; **(E)** the characteristic correlation length $\xi$, determined by fitting each $g(r)$ curve to an exponential function (color-coded dashed lines), and denoted by the corresponding intersection of the horizontal dashed line at $g = e^{-1}$; **(F)** the cluster size at CDF = 0.5, i.e., the median cluster size, as denoted by the intersection of the horizontal dashed line with each CDF. The data shown in D-F are from images collected at 1 h for pS (black) and pT (blue) and 28 hrs for pS (cyan). Further details regarding the analyses depicted in D-F are described in Methods and SI Methods.
9

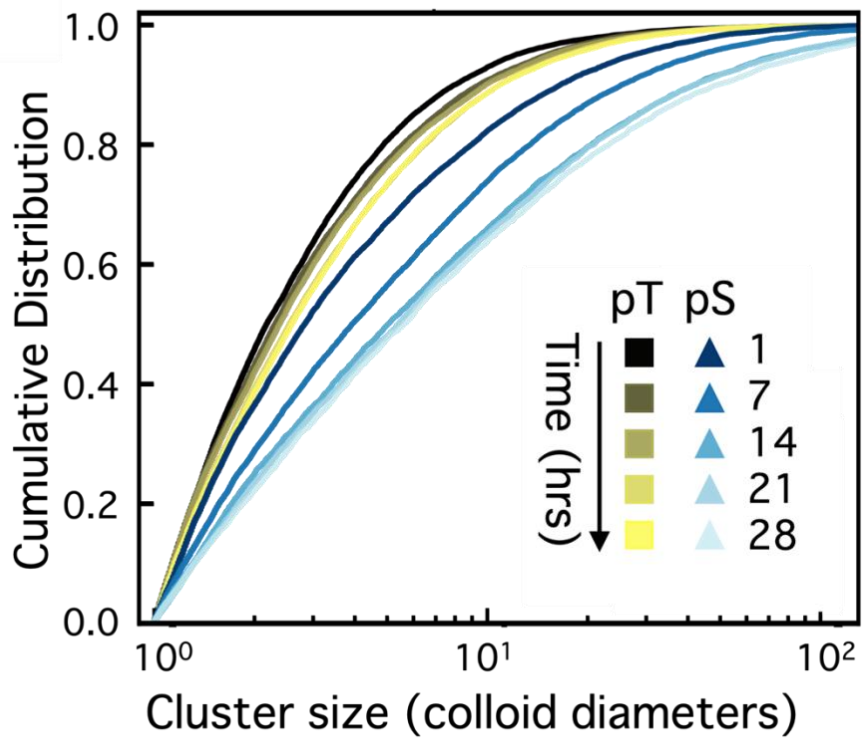

**Figure S4. Time-dependence of KaiC-mediated cluster formation quantified via cluster size analysis. (A)** CDFs of cluster sizes for 5 different times between 1 hr and 28 hr for pT (yellow) and pS (cyan) with color shade indicating time according to the legend. CDFs are used to determine median cluster sizes plotted in Fig 3, as depicted in Fig S3F.

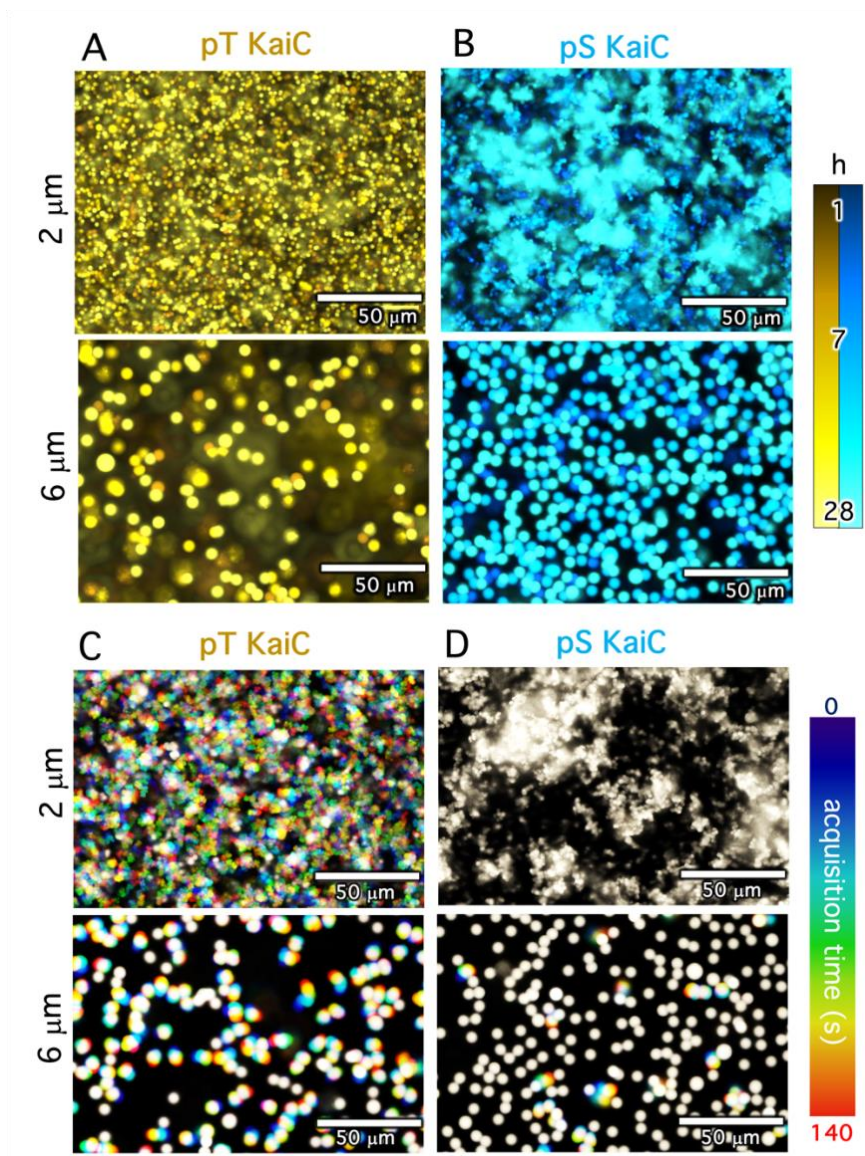

**Figure S5. Clock-mediated crosslinking of 2 µm and 6 µm colloids display the same self-assembly kinetics as 1 µm colloids in the presence of biotinylated Kai proteins.** (A,B) Colorized temporal overlays of fluorescence microscopy images taken 1 h, 7h, and 28 h (dark to light) after mixing 2 µm (top) and 6 µm (bottom) colloids with pT KaiC (A, yellow) and pS KaiC (B, cyan) show cluster growth over time with pS that is absent for pT. (C, D) Temporal color-coded collapses of 140-s videos taken at 28 h for 2 µm (top) and 6 µm (bottom) colloids in the presence of pT (C) or pS (D). White indicates lack of motion due to the exact overlay of all colors corresponding to different frames. Colloids in the presence of pT maintain mobility, seen as more color in the images, while pS induces aggregation and crosslinking that stalls colloid motion. The concentrations of colloids, proteins and reagents, as well as imaging parameters, are identical to those for the 1 µm colloid experiments shown in Figs 2-4. As shown, timed aggregation, dependent on the phosphorylation state of KaiC is preserved for different sizes of colloids.

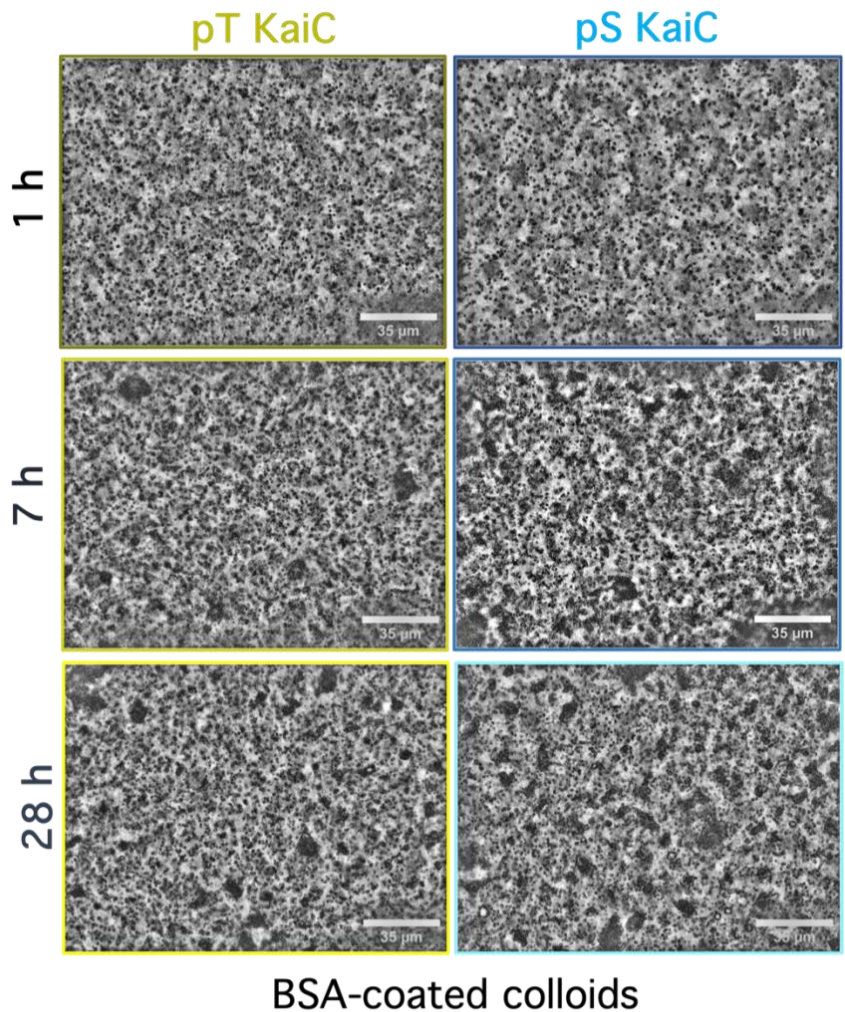

**Figure S6. Brightfield microscopy of suspensions of passivated colloids incubated with pT KaiC (left) or pS KaiC (right) with 55% b-KaiB.** Brightfield microscopy images of colloids of the same size and material (1-µm, polystyrene, Polysciences, Inc) as streptavidin-coated colloids used in the Kai-colloid crosslinking experiments but coated with BSA rather than streptavidin. Images show minimal clustering and aggregation for both pT and pS over the course of 28 hrs. The modest clumping observed at later times is not specific to the type of KaiC, demonstrating that the Kai-specific timed self-assembly of colloids that we observe in experiments is due to specific interactions between KaiC and colloid-bound KaiB proteins. KaiC concentration in all experiments is 6.67 µM, the highest concentration used in experiments in the main text. Imaging parameters are the same as in the main text but brightfield instead of fluorescence is used because the colloids are not fluorescently-labeled.

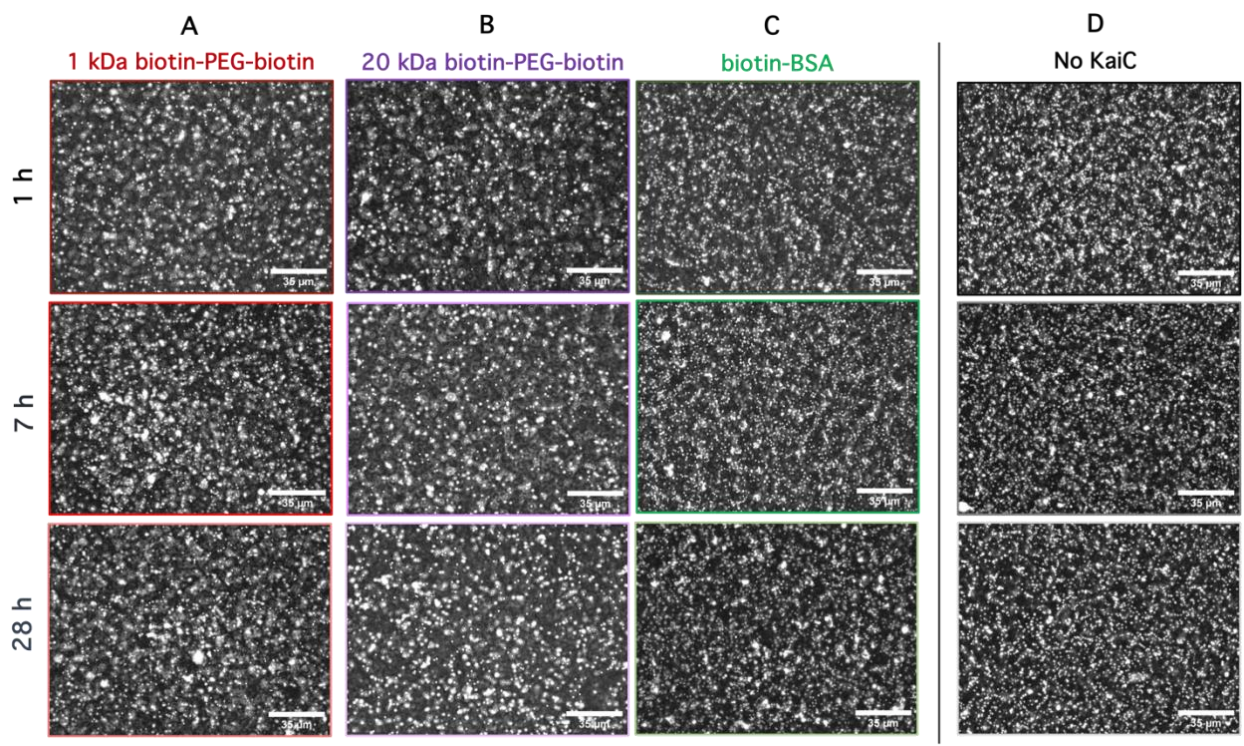

**Figure S7. Streptavidin-coated colloids in the presence of biotinylated PEG or BSA (no Kai proteins), or in the presence of biotin-KaiB and WT KaiB but no KaiC, show no apparent crosslinking or clustering.** Suspensions of streptavidin-coated colloids, with identical conditions to those shown in the main text, but with (**A-C**) Kai proteins replaced with different biotinylated constructs that could, in principle, crosslink streptavidin-coated colloids. To demonstrate that the timed crosslinking we observe is specific to the Kai clock proteins, we replaced Kai proteins with different constructs that have multiple biotins to allow for potential crosslinking of streptavidin-coated colloids: (**A**) 1 kDa biotin-PEG-biotin (PG2-BN-20k, Nanocs Inc), (**B**) 20 kDa biotin-PEG-biotin (PG2-BN-1k, Nanocs Inc), and (**C**) biotinylated BSA, with 8-16 biotins per BSA (A8549, Sigma-Aldrich). Both PEGs have biotin labels on either end and only differ in the lengths of the PEG linker (1 vs 20 kDa). BSA has 8-16 randomly distributed biotins per BSA molecule. In the images shown, the molarity of the constructs was matched to the highest KaiC molarity used in Kai-crosslinking experiments. As shown, no discernible crosslinking or clustering is observed over the course of 28 hrs for all biotinylated linkers. (D) Experiments performed identically to those in the main text but in the absence of KaiC (only biotin-KaiB and WT KaiB) show no discernible crosslinking, demonstrating that the complexation of KaiC with KaiB is essential to colloidal self-assembly. Top and bottom row images are the same as in Fig 2F,G.

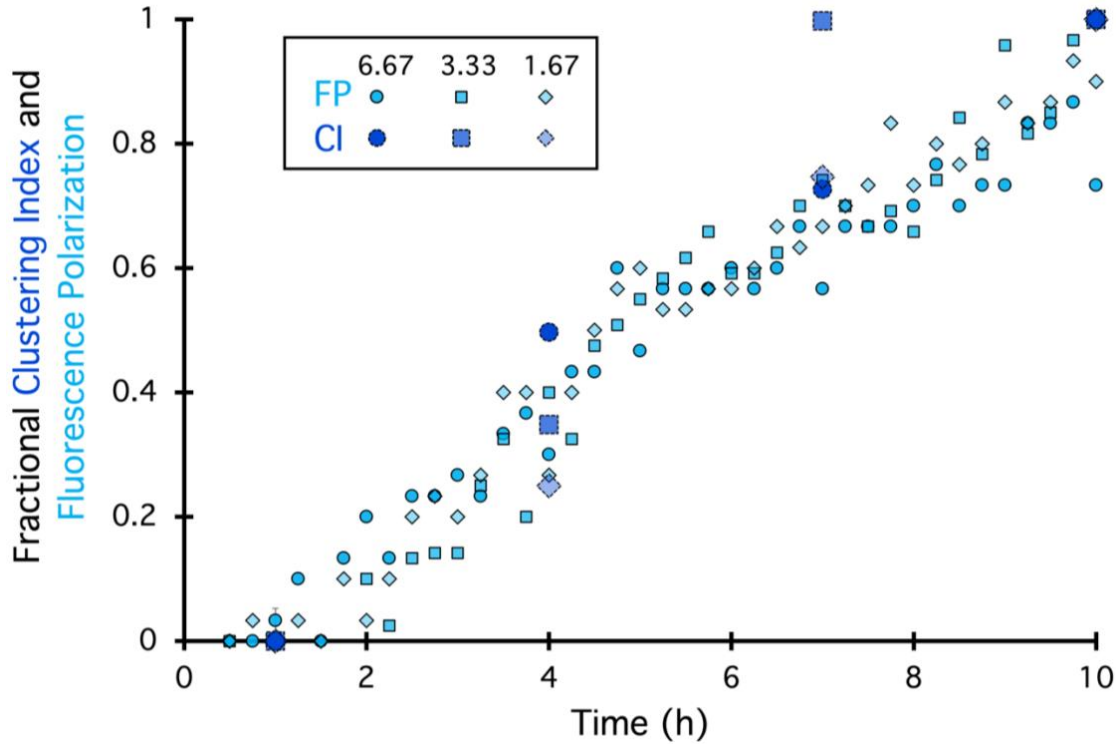

**Figure S8. The time constant for colloidal crosslinking is robust to the concentration and mirrors the time constant of KaiBC complex formation.** Time-dependent fractional clustering index (CI, blue, dashed borders), determined from the SIA correlation length $\xi$ as described in the main text, for 1 µm colloids in the presence of varying concentrations of pS KaiC: 6.67 µM (circles), 3.33 µM (squares) and 1.67 µM (diamonds). Fluorescence polarization (FP, cyan, solid borders), an indicator of KaiBC complex formation, measured for the same Kai concentrations as the colloid experiments, is also plotted using smaller symbols. These data demonstrate that the rate of the slow self-assembly of colloids in the presence of pT KaiC is independent of Kai concentration, as is the rate of KaiBC complex formation. Moreover, the time-courses of CI and FP data are highly similar, a strong indicator that the time constant for colloidal self-assembly is controlled by the rate of KaiBC assembly.

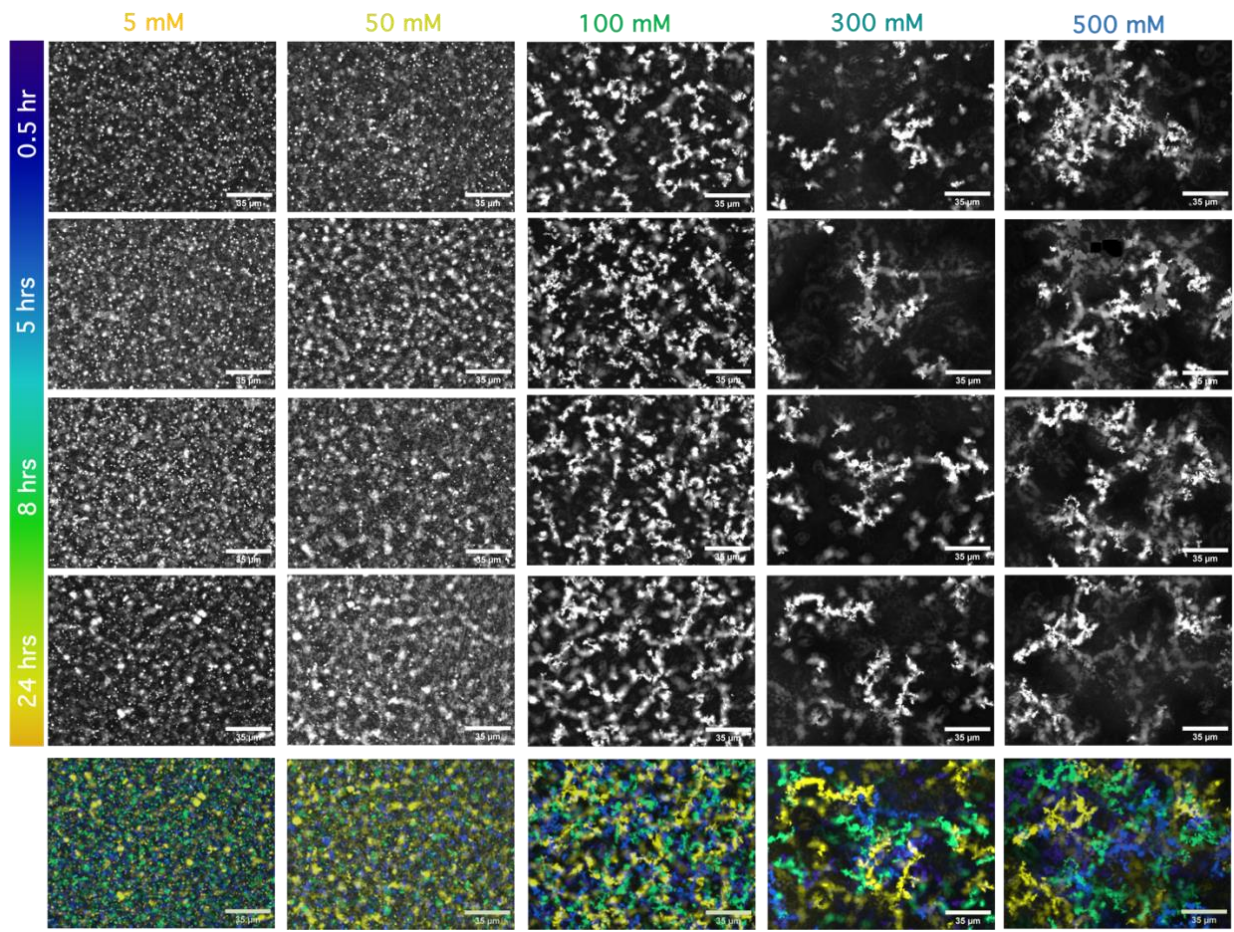

**Figure S9. Both timing and extent of salt-mediated clustering of colloids depend on salt concentration.** Streptavidin-coated colloids in the same buffer conditions as Kai-crosslinking experiments but in the absence of Kai proteins and with varying added concentrations of $MgCl_2$, listed from left to right. Images were collected 0.5 hr (top), 5 hrs (row 2), 8 hrs (row 3) and 24 hrs (row 4) after adding $MgCl_2$. Bottom row images are overlays of images from all 4 time points, color-coded according to time based on the color scale to the left. Increasing $[MgCl_2]$ causes increasing degrees of aggregation and clustering. The timing of the onset and saturation of aggregation also varies with $[MgCl_2]$. At 5 mM, only small-scale clustering is observed starting at ~8 hrs and continuing to increase without saturating up to 24 hrs. At 50 mM, more clustering is observed and the onset is earlier (~5 hrs) than with 5 mM, but no saturation is apparent. For $[MgCl_2] \geq 100$ mM, the onset of clustering is immediate and appears to saturate at ~5 hrs (100 mM) or sooner (300 mM, 500 mM). The timing of clustering appears to be directly linked to the salt concentration, unlike the timing of clock-mediated crosslinking.

**Movie S1. Animations of simulations of colloids crosslinking under the action of (A) no crosslinkers (N), (B) permanent crosslinkers (P), and in cases in which each connection between two colloids is formed by (C) 1, (D) 2 or (E) 3 oscillatory crosslinkers**. Animations show the system dynamics over the course of 48 hrs with 1 μm colloids shown as red circles and bonds shown as purple connecting lines. The timestamp in the bottom left lists time in hours. Images depict a 2D square box of length 50 μm in which the colloids are confined.

(A) [No Bond](No Bond)
(B) [Permanent Bond](Permanent Bond)
(C) [1 Oscillatory Bond](1 Oscillatory Bond)
(D) [2 Oscillatory Bonds](2 Oscillatory Bonds)
(E) [3 Oscillatory Bonds](3 Oscillatory Bonds)

**Table S1. Table of parameters used in simulations.** Key parameters and their values used in the model and simulations. Numerical estimates of all parameters are based on experimental values as described in the Source/Rationale column.

| Simulation Parameters | Value | Source/Rationale |
|---|---|---|
| 2D colloid packing fraction, $p_f$ | 16% | Area fraction of pixels above threshold in binarized images of pT colloids |
| Simulation unit lengthscale, $\sigma$ | 1 μm | Diameter of colloidal particle |
| Simulation unit timescale, $\tau$ | 2.5 s | Time for colloid to diffuse across its diameter in water, $\tau = \sigma^2/D$ |
| Crosslinker oscillation period, $T$ | $34560\tau$ | ~24 hrs, set to match KaiABC period |
| Brownian dynamics timestep, $dt$ | $0.001\tau$ | Ensures numerically stability |
| Characteristic timescale of KaiB-KaiC interaction, $\tau_F$ | $4 \times 10^4$ s | Obtained by fitting experimental FP data for pS (Fig 4) to an exponential |
| Probability of colloid crosslinking, $p_0$ | 6e-5 | $p_0 \simeq 1 - e^{-\tau/\tau_F} \approx \tau/\tau_F$ |
| Equilibrium distance between centers of crosslinked colloids, $r_0$ | $1.01\sigma$ | Sum of colloid diameter and ~10 nm size of KaiBC complex |
| Spring constant of crosslinks, $K$ | $9 k_B T/\sigma$ | Phenomenological estimate |